\documentclass[a4paper,11pt]{article}
\pdfoutput=1 

\usepackage{jcappub}
\usepackage{aas_macros} 
\usepackage{fontawesome}
\usepackage[T1]{fontenc}
\usepackage{amsmath,bm}

\newcommand{\bs}{\boldsymbol}
\newcommand{\fNL}{f_{\rm NL}}

\newcommand{\dd}{\mathrm{d}}
\usepackage{hyperref}
\usepackage{url} 
\usepackage{graphicx}
\usepackage{subcaption}
\usepackage{comment}

\title{\boldmath Multipoles of the galaxy bispectrum on a light cone: wide-separation and relativistic corrections}
\author[a,1]{Chris Addis,\note{Corresponding author.}}
\author[b,a]{Caroline Guandalin,}
\author[a,c]{Chris Clarkson}

\affiliation[a]{Astronomy Unit, School of Physical \& Chemical Sciences, Queen Mary University of London,  London E1 4NS, UK}
\affiliation[b]{Institute for Astronomy, University of Edinburgh, Royal Observatory, Blackford Hill, Edinburgh EH9 3HJ, UK}
\affiliation[c]{Department of Physics \& Astronomy, University of Western Cape, Cape Town 7535, South Africa}

\emailAdd{c.l.j.addis@qmul.ac.uk}
\emailAdd{caroline.guandalin@roe.ac.uk}
\emailAdd{chris.clarkson@qmul.ac.uk}

\abstract{The galaxy bispectrum provides access to correlations among different scales that cannot be captured by the power spectrum alone, and with the Stage-IV galaxy surveys it enables the possibility of detecting both primordial non-Gaussianity (PNG) and general relativistic effects. Accounting for wide-separation corrections, which arise from the loss of symmetry in the correlation of widely separated points on the past light cone, is essential for their accurate modelling and detection. These corrections can be included perturbatively to the standard bispectrum and we compute them analytically for a generalised line of sight, including the radial evolution contribution to the bispectrum for the first time. We show that the first-order corrections entering the odd multipoles with respect to the line of sight are large, up to $10 \%$ of the bispectrum monopole, and need to be included when considering the leading-order relativistic effects that could be detectable with surveys like DESI and Euclid. The second-order wide-separation and relativistic contributions, including their mixing terms, enter into the even multipoles and therefore have implications for analysis of PNG and we show, for the local type, they can mimic $f_{\rm NL}$ of order 10 in the squeezed limit. We present full analytic expressions for all these contributions to the local bispectrum and its multipoles which are implemented in a new publicly available \textit{Python} package \textsc{CosmoWAP}.}

\begin{document}

\maketitle

\flushbottom

\section{Introduction}
Much effort has been employed in the use of summary statistics of galaxy clustering in redshift space to test the theoretical assumptions entering our cosmological models, from the initial conditions that gave origin to the large-scale structure (LSS) of the universe, up to the theory of gravity responsible for structure formation.

If the initial conditions were perfectly Gaussian and gravitational collapse was a linear process, the mean of the density field and its power spectrum would provide a full statistical description of the field. However, the nonlinear dynamics correlates different scales and higher-order moments are thus required to better characterise the statistical distribution of the observed fields. The bispectrum -- the lowest higher-order statistics beyond the power spectrum -- has been used to quantify departures from the Gaussian assumption in both matter \citep[see][]{scoccimarro2001,feldman2001} and temperature fields \citep[e.g.,][]{heavens1998,planck2018PNG}. The latter gives the tightest constraints to date on the level of non-Gaussianities in the initial conditions due to the small level of nonlinear structure formation at the redshifts of the cosmic microwave background (CMB). However, as CMB experiments push the limits of angular resolution to detect temperature fluctuations at smaller scales, the number of available modes are inevitably being exhausted. On the other hand, galaxy redshift surveys probe the 3D distribution of matter, resulting in many more modes to constrain cosmological parameters. 

Over the past decades, these surveys have grown in the number density of detected objects and also in area and redshift range coverage \citep[see, for example, Figure 1 of][]{schlegel2022}. In particular, we are witnessing now the emergence of the so-called Stage-IV spectroscopic galaxy surveys, with Dark Energy Spectroscopic Instrument (DESI) \cite{desi} and Euclid \cite{euclid} already in operation, and others such as SPHEREx \cite{Dore_SPHEREx} and Nancy Grace Roman Space Telescope (WFIRST) \cite{wfirst} to start in the upcoming years. They will enable the study of primordial physics \cite{sailer2021} and tests of general relativity \cite{beutler2020} with high accuracy due to the large sky fraction mapped by these surveys ($f_{\rm sky}\gtrsim 0.3$ \cite{sailer2021}). In addition to these surveys, the Square Kilometre Array Observatory (SKAO) will conduct a low-redshift neutral hydrogen (HI) galaxy survey, during Phase-1 (SKAO1) \cite{bacon2018}, with spectroscopic precision by mapping the 21-cm radio emission of galaxies. Finally, the conceptual Phase-2 (SKAO2) would expand the observations to higher redshifts, and over a larger area (e.g., \cite{Bull_2016}). 

In order to draw robust conclusions from these surveys, the observables need to be corrected for projection effects that emerge from the fact that observations are made on our past light cone \cite{Bonvin_2011}. The most prominent and well-studied effect are the standard redshift-space distortions (RSD) \cite{kaiser1987}, which are manifested as anisotropies due to the projection of the peculiar velocity of galaxies projected along the line-of-sight (LOS) direction. However, other effects that appear in the observed redshift, such as gravitational redshift, Doppler corrections and other relativistic effects \cite{yoo2009} are also relevant when we constrain cosmology with the galaxy number counts (e.g., see \cite{yoo2015}).

Analyses involving redshift-space quantities typically decompose the induced LOS dependence of the statistics into multipoles with respect to its orientation to the LOS. In this so-called redshift space, the bispectrum has been shown to break degeneracies between the linear bias of the tracers $b_1$, the growth rate of structures $f$, the amplitude of linear perturbations $\sigma_8$ \cite{scoccimarro1999,gil2014}, and neutrino masses \cite{hahn2020}, for example. Finally, further interest is directed towards the relativistic projection effects, whose leading-order contribution generates odd multipole moments in the galaxy bispectrum \cite{Clarkson_2019,deWeerd_2019,jeong2020} and which can act as a direct probe of gravity on cosmological scales.

The convention for these multipoles is that they are calculated from the $n$-point function at local regions (defined by a singular LOS) where statistical homogeneity is assumed \cite{Scoccimarro_2015}. This approximation breaks down at large scales, since the statistics of the density field at each point in the local correlation function is different. Corrections to this singular LOS approximation can be done through a series expansion, where it becomes dependent on the actual choice of the LOS in the triplet \footnote{Alternatively one could decompose the bispectrum in a Spherical Fourier-Bessel basis which naturally fully includes this information, though at the cost of significant computational complexity - see \cite{Benabou_2024} for recent work with the bispectrum.}.

Although the angular part of these corrections, induced as we leave the local plane-parallel limit, has been extensively studied for the case of the two-point statistics \cite{szalay1998,matsubara2000,szapudi2004,papai2008,bertacca2012,montanari2012,reimberg2016,castorina2018,beutler2019,beutler2020,Paul_2022,Joliceur_2024}, giving rise to the so-called wide-angle corrections, their impact on the galaxy bispectrum \cite[for example, see][]{bertacca2018} and corrections that appear by breaking statistical homogeneity radially \cite{Bonvin_2014,reimberg2016,beutler2020,Paul_2022,Joliceur_2024} are much less common. It was only recently that wide-angle corrections were computed for the bispectrum in a Cartesian-Fourier basis in \cite{Noorikuhani_2022}. The consistency of \cite{Noorikuhani_2022} was then verified with a different basis expansion in \cite{Pardede_2023}. In these works, it is shown that wide-angle effects in the bispectrum can bias the detection of PNG of the local type, a specific signal which emerges from a quadratic correction to the primordial gravitational potential \cite{komatsu2001,Dalal_2008}.

Indeed, with Stage-IV surveys, these corrections become relevant not only due to the increased precision, but also as these surveys map increased volume the scales that are correlated increase. In this work, we attempt to provide a consistent picture of the relative contribution of relativistic, wide-angle and radial evolution terms to the bispectrum multipoles over a range of bispectrum shapes and scales and how this depends on the LOS used to describe the local triplet. 

Wide-separation terms are suppressed by a factor $(1/k \, d)^m$ for a given order in the perturbative expansion, while the relativistic terms have suppression factors of $(\mathcal{H}/k)^n$. In this work, we calculate terms up to second-order ($m+n \leq2$), including cross-terms, and we highlight how wide-separation corrections have the potential to impact both the analysis of both the imaginary relativistic contribution and also PNG. 

This paper is organised as follows. In Section \ref{sec:local}, we introduce the local bispectrum along with the relevant perturbation theory, as well the relativistic corrections we consider. In Section \ref{sec:WS_corrections}, we present the perturbative framework to include wide-separation effects in the bispectrum. The results are presented and discussed in Section \ref{sec:results}, and we conclude in Section \ref{sec:conclusions}. 

\begin{figure}[tbp]
    \centering 
    \includegraphics[width=.55\textwidth]{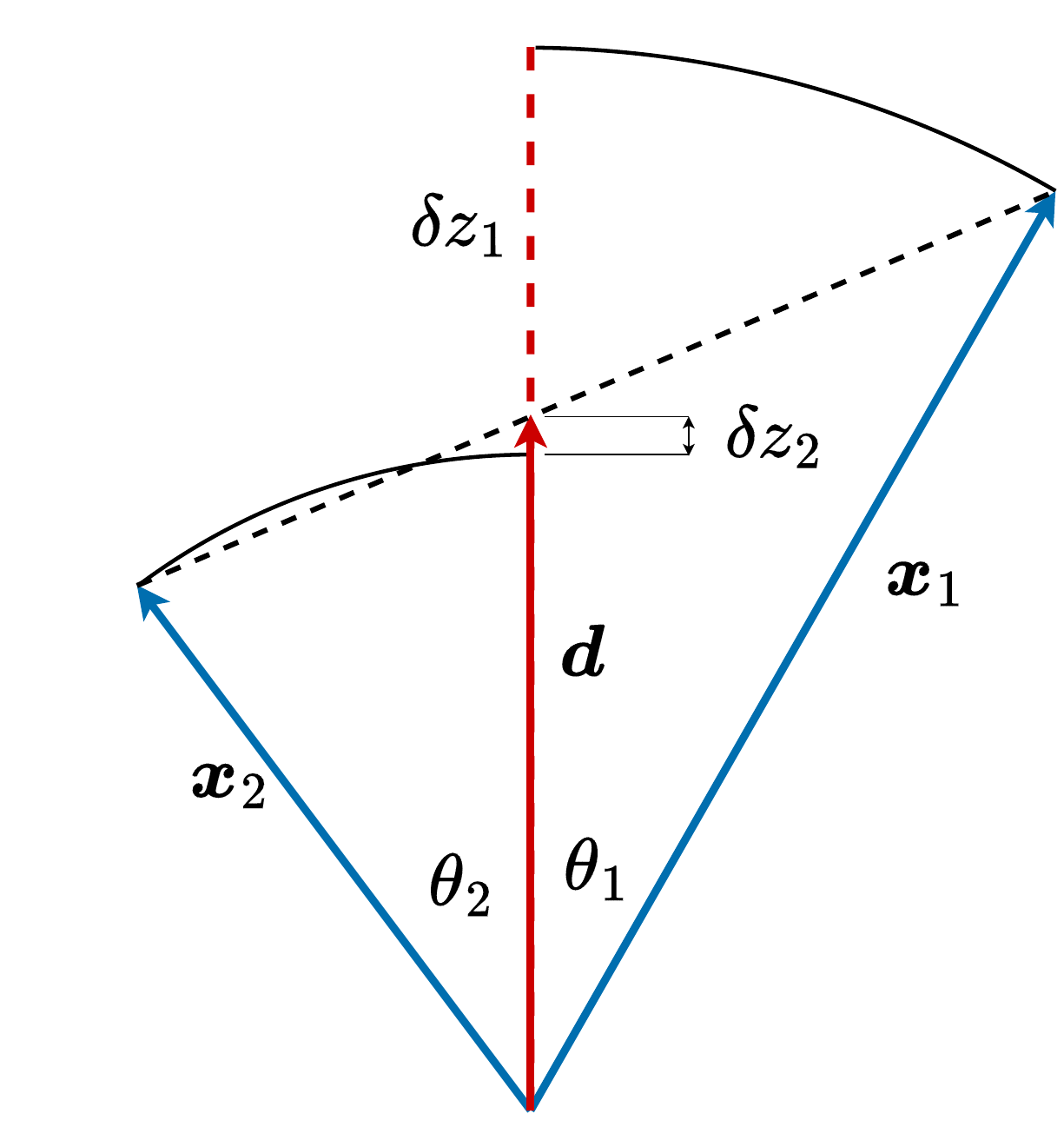}
    \caption{\label{fig:real_space_pk} Real space two-point function, where deviations from the plane-parallel and constant redshift limit are respectively expressed in terms of $\theta_1,\theta_2$ and $\delta z_1,\delta z_2$ for a given LOS $\bs{d}$.}
\end{figure}

\section{The Local Bispectrum}\label{sec:local}

    In the case of a statistically homogeneous field our statistics are more straightforward with fewer degrees of freedom, e.g. in the case of the power spectrum it is diagonal and the bispectrum is zero for open triangles. Indeed, without statistical homogeneity the standard approach to expand the Fourier space correlators into multipoles with respect to the orientation to the LOS becomes ill-defined \cite{Zaroubi_1993, hamilton1998,castorina2018} and different statistics are needed. 

    While we assume the density field to be statistically homogeneous in real-space for a constant time slice, in reality, when we observe on a light cone, this symmetry is broken by a couple of effects. Firstly, redshift-space distortions cause our field in redshift space to depend on the LOS; for a survey outside the plane-parallel limit, different regions within the survey will be statistically different due to the differing LOS. Secondly, on the past light cone, the statistics change radially as the density field evolves with redshift.

    Therefore, it is useful and conventional to define local regions which are approximated to be statistically homogeneous \cite{Scoccimarro_2015}, and thus can be described by a singular LOS. In this case, therefore, the statistics are diagonal. 
    
    The local 3-point function is thus described by a triplet of points defined at a certain point in space, which define our local bispectrum after a Fourier transformation,
    \begin{equation}
        \begin{aligned}\label{eq:local_bispectrum}
            B^{\rm loc}(\bs{k}_{1},\bs{k}_{2},\bs{d}) \equiv \int \dd^3 \bs{x}_{13}\dd^3 \bs{x}_{23} \,{\rm e}^{-i(\bs{k}_1 \cdot \bs{x}_{13} + \bs{k}_2 \cdot \bs{x}_{23})}\,\langle\delta(\bs{x}_1)\delta(\bs{x}_2)\delta(\bs{x}_3)\rangle, 
        \end{aligned}
    \end{equation} 
    where $\bs{x}_{13} = \bs{x}_{1} - \, \bs{x}_{3}$ and the dependence on $\bs{d}$ is implicit in the 3-point correlation function given by $\langle\delta(\bs{x}_1) \delta(\bs{x}_2) \delta(\bs{x}_3 )\rangle \equiv \zeta(\bs{x}_1,\bs{x}_2,\bs{x}_3) \equiv \zeta(\bs{x}_{13},\bs{x}_{23},\bs{d})$.

    The full local bispectrum in redshift space can parameterised with six degrees of freedom: $B_{\rm loc}(q_1,q_2,q_3,\mu_1,\phi_{12};d)$, three of which, $k_1,k_2,k_3$, describe the triangle shape and two more, $\theta_1, \phi_{12}$, describe the orientation of the triangle with respect to the LOS, and $d$ is the separation of the observer to the triplet. We define $\theta$ as the angle between $\bs{k}_1$ and the LOS, $\bs{d}$, such that $\cos(\theta_1)=\mu_1= \bs{k}_1\cdot\bs{d}$ and $\phi$ are the azimuthal angles between $\bs{k}_2$ and $\hat{\bs{d}}$ in the plane normal to $\hat{\bs{k}}_1$. Therefore, they satisfy 
    \begin{equation}
    \mu_2 = \mu_1\cos{\theta} + \sqrt{1-\mu^2}\sin{\theta}\cos{\phi_{12}},
    \end{equation}
    where $\cos{\theta}= \hat{\bs{k}}_1 \cdot \hat{\bs{k}}_2$. 

    \subsection{Multipoles estimator}
    
    If we construct the standard `Scoccimarro' estimator for the bispectrum multipoles \cite{Scoccimarro_2015} 
    \begin{equation}\label{eq:estimator}
    \begin{aligned}
        \hat{B}_{\ell m}(k_1,k_2,k_3) = \frac{k_f^{3}}{N^{T}_{123}}\int_{\mathcal{S}_1}& \frac{\dd^3 \bs{q}_1}{(2 \pi)^3} \int_{\mathcal{S}_2} \frac{\dd^3 \bs{q}_2}{(2 \pi)^3} \int_{\mathcal{S}_3} \frac{\dd^3 \bs{q}_3}{(2 \pi)^3} \,\,\delta^D(\bs{q}_{123})  \int  \dd^3 \bs{x}_{1}\, \dd^3 \bs{x}_{2}\, \dd^3 \bs{x}_{3}\\ &\times {\rm e}^{-i(\bs{k}_1 \cdot \bs{x}_{13} + \bs{k}_2 \cdot \bs{x}_{23})}\,\,\delta_W(\bs{x}_1)\,\delta_W(\bs{x}_2)\,\delta_W(\bs{x}_3) \, Y^*_{\ell,m}(\hat{\bs{q}}_1 \cdot \hat{\bs{d}},\phi),
    \end{aligned}
    \end{equation}
    where the orientation of the triangle relative to the LOS $\hat{\bs{d}}$ is decomposed into spherical harmonics about $\hat{\bs{d}}$. Note, however, the $m\neq0$ parts are not computable with separable Fast Fourier Transforms (FFT), and so it is common to just focus on the $m=0$ part, which can be written in terms of a Legendre basis. Further, since this is measured in a survey, we define the windowed density field $\delta_W(\bs{x})= W(\bs{x})\delta(\bs{x})$, where $W(\bs{x})$ is the survey window function. The Fourier space integrals are discrete sums $\int_{\mathcal{S}_i} \dd^3 \bs{q}_i F(\bs{q}_i) \equiv \sum_{\mathcal{S}_i} F(\bs{q}_i)$ over thin $k$-space shells $\mathcal{S}_i$ of width $\Delta k$, centred at $k_i$, such that $\mathcal{S}_i \equiv \mathcal{S}(k_i|\Delta k)$ is the region of $k$-magnitudes contained by a given $k_i$-bin, $k_i -  \Delta k/2 \leq k \leq k_i + \Delta  k/2$. Here, $V_{123}$ represents the number of closed triangles formed from the triplet of bins,
    \begin{equation}\label{number of triangles}
        V_{123} = k_f^{-3}\int_{\mathcal{S}_1} \dd^3 \bs{q}_1 \int_{\mathcal{S}_2} \dd^3 \bs{q}_2 \int_{\mathcal{S}_3} \dd^3 \bs{q}_3 \,\,\delta^D(\bs{q}_{123}),
    \end{equation}
    where $k_f$ is the fundamental frequency in Fourier space. We can then construct the equivalent unwindowed\footnote{We discuss the impact of the survey window on the bispectrum in the context of wide-separation effects in Appendix \ref{ap:window_convolution}} theoretical expectation that corresponds to the `Scoccimarro' estimator for each multipole $B_{\ell m}$ \cite{Scoccimarro_2015}. In the continuous limit, using the change in coordinates
    \begin{equation}
        \int \dd^3 \bs{x}_{1} \dd^3 \bs{x}_{2} \dd^3 \bs{x}_{3} \rightarrow \int \dd^3 \bs{d} \int \dd^3 \bs{x}_{13} \dd^3 \bs{x}_{23}, 
    \end{equation}
    and splitting the integration in each shell
    \begin{equation}
        \int_{\mathcal{S}_i} \dd^3 \bs{q}_i = \int^{k_i + \Delta  k/2}_{k_i - \Delta  k/2} \dd q_i q_i^2\int\dd\Omega_{q_i},
    \end{equation}
    where $\int\dd\Omega_q$ represents an integral over solid angle, the corresponding theoretical multipoles can be expressed in the familiar form:
    \begin{equation}
    \begin{aligned}\label{eq:bk_multipole}
        B_{\ell m}(k_1,k_2,k_3) = \int \frac{\dd^3 \bs{d}}{V_s}\left[\int \frac{\dd \Omega_q}{4 \pi}B^{\rm loc}(q_1,q_2,q_3,\mu_1,\phi_{12};d) Y^*_{\ell m}(\mu_1,\phi_{12})\right].
    \end{aligned}
    \end{equation}
    with $V_s$ representing the normalisation over the survey volume. It is convenient to define the term in the square bracket above as the local bispectrum multipole, such that the multipoles measured with a given survey corresponds to an average of this local quantity over the entire survey volume.
    
    Equivalently, we could consider alternative decompositions for the anisotropic local bispectrum, such as the tri-polar spherical harmonic (TripoSH) \cite{Sugiyama_2018} or other modal approaches (e.g., see \cite{Byun_2022}), where wide-separation corrections are still computed from the local bispectrum, but they will enter a different set of multipoles and modes.

\subsection{Perturbation theory}
In this section we give a brief overview of the perturbation theory that forms the basis of our later calculations; the aim is to write the theoretical local bispectrum with explicit dependence on the `end-point' positions, $\bs{x}_1,\bs{x}_2,\bs{x}_3$, including where we have redshift dependent functions. See \cite{Bernardeau_2001} for a more detailed review of these topics. 

The local bispectrum correlates density fields at different radial comoving distances, $x_i$, which corresponds to different redshifts, $z_i$. Therefore, we express this redshift dependence in terms of the radial comoving distances. We also adopt the following notation for real and Fourier integrals;
\begin{equation}
    \int_{\bs{x}} \equiv \int \dd^3 \bs{x} \, \, , \, \, \int_{\bs{q}} \equiv \int \frac{\dd^3 \bs{q}}{(2 \pi)^3}.
\end{equation}

In standard perturbation theory we can express both the matter overdensity $\delta(\bs{x})$ and velocity divergence $\theta(\bs{x})= \nabla \cdot \bs{v}(\bs{x})$ fields neatly in Fourier space as a perturbative sum where each $n^{\rm th}$ order field can be expressed in terms of $n$ products of first order density field with some coupling kernel. For example, we can write

\begin{subequations}\label{eq:overdensityPT}
\begin{align}
\delta(\bs{k},x) &= \sum_{n=1}^{\infty} \int_{\bs{k}_1,...,\bs{k}_{n-1},\bs{k}_n}\delta_{D}(\bs{k}-\bs{k}_1-...-\bs{k}_n)F_n(\bs{k}_1,...,\bs{k}_n,x)\delta^{(1)}(\bs{k}_1,x)...\delta^{(1)}(\bs{k}_n,x),
\\
\theta(\bs{k},x) &= -f\mathcal{H}\sum_{n=1}^{\infty} \int_{\bs{k}_1,...,\bs{k}_{n-1},\bs{k}_n}\delta_{D}(\bs{k}-\bs{k}_1-...-\bs{k}_n)G_n(\bs{k}_1,...,\bs{k}_n,x)\delta^{(1)}(\bs{k}_1,x)...\delta^{(1)}(\bs{k}_n,x),
\end{align}
\end{subequations}
where $F_n(\bs{k}_1,...,\bs{k}_n,x_1)$ and $G_n(\bs{k}_1,...,\bs{k}_n,x_1)$ are the perturbation theory kernels responsible for the coupling of scales, $f$ is the linear growth rate of matter perturbations, and $\mathcal{H}$ is the conformal Hubble parameter.

At first order, the density field can be separated into its temporal and spatial parts:
\begin{equation}
    \delta^{(1)}(\bs{k}_1,x_1) = D(x_1) \delta^{(1)}(\bs{k}_1),
\end{equation}
where the time evolution is parameterised by the linear growth function, $D(x_1)$, satisfying the equation  
\begin{equation}
    \mathcal{H}^2(x_1)(1+z(x_1))^2 D_{zz}(x_1)+[\mathcal{H}_z(x_1)\mathcal{H}(x_1)(1+z(x_1))^2] D_z=\frac{3 H_0^2 \Omega_{m,0}}{2 a(x_1)} D(x_1),
\end{equation}
where derivatives of a function $F$ with respect to redshift are denoted by $F_z$.

For the case of a field at second order, this can be expressed in terms of a convolution of two first-order fields with the relevant coupling kernel (e.g., Equation~\ref{eq:overdensityPT}). At second order, this split is not so trivial; however, it can be solved to provide a solution for the Fourier-space Newtonian coupling kernels (see, for example, \cite{Matsubara_1995,Tram_2016}):
\begin{subequations}\label{eq:f2g2kernels}
\begin{align}
    F_2(\bs{q}_1,\bs{q}_2,x_1) &= \frac{1}{2}[1 + K(x_1)] + \frac{1}{2}\hat{\bs{q}}_1 \cdot \hat{\bs{q}}_2 \left(\frac{q_2}{q_1}+\frac{q_2}{q_2}\right)+ \frac{[1-K(x_1)]}{2} (\hat{\bs{q}}_1 \cdot \hat{\bs{q}}_2)^2,
    \\
    G_2(\bs{q}_1,\bs{q}_2,x_1) &=  C(x_1)+ \frac{1}{2}\hat{\bs{q}}_1 \cdot \hat{\bs{q}}_2 \left(\frac{q_2}{q_1}+\frac{q_2}{q_2}\right)+ [1 - C(x_1)](\hat{\bs{q}}_1 \cdot \hat{\bs{q}}_2)^2,
\end{align}
\end{subequations}
where, following the same notation as \cite{Matsubara_1995}, the redshift-dependent coefficients are defined by the relationship of the first and second-order growth functions, $D$ and $F$ respectively, such that
\begin{equation}
K(x_1) \equiv \frac{F(x_1)}{D^2(x_1)}, \, \, C(x_1) \equiv \frac{F'(x_1)}{ 2D(x_1) D'(x_1)}.
\end{equation}

The second-order growth rate satisfies a differential equation similar to the one at first-order, but with an additional source term
\begin{equation}
    \mathcal{H}^2(x_1)[1+z(x_1)]^2 F_{zz}(x_1)+\left(\mathcal{H}_{z}(x_1)\,\mathcal{H}(x_1)\,[1+z(x_1)]^2\right) F_{z}(x_1)=\frac{3 H_0^2 \Omega_{m,0}}{2 a(x_1)} [F(x_1)+D^2(x_1)].
\end{equation}
In the Einstein-de-Sitter limit, we recover the standard approximation for the second order kernels, where $K(x_1) = C(x_1) = 3/7$.

The standard first- and second-order redshift space Newtonian kernels with explicit $\bs{x}_1,\bs{x}_2$ and $\bs{x}_3$ are given by \cite{scoccimarro1999,Verde_1998}
\begin{equation}
        Z^{(1)}_{\rm N}(\bs{q},\bs{x}_1) = D(x_1)[b_1 + f (\hat{\bs{q}} \cdot \hat{\bs{x}}_1)^2],
\end{equation}
and
\begin{equation}
\begin{aligned}
    Z^{(2)}_{\rm N} = \, & D^2(x_1)   \left(b_1(x_1) \left(F_2(\bs{q}_1,\bs{q}_2,x_1)\right) +f(x_1)\frac{(\bs{q}_{12}\cdot \hat{\bs{x}}_1)^2}{q^2_{12}}\left( G_2(\bs{q}_1,\bs{q}_2,x_1)\right) \right. \\ & \left.+\frac{b_2(x_1)}{2} \, +b_{\Gamma_2}(x_1)S(\bs{q}_1,\bs{q}_2)\right. \\ & \left. + \frac{f(x_1)}{2}(\bs{q}_{12}\cdot \hat{\bs{x}}_1)\left[b_1(x_1)\left(\frac{(\bs{q}_{1}\cdot \hat{\bs{x}}_1)}{q_1^2}+\frac{(\bs{q}_{2}\cdot \hat{\bs{x}}_1)}{q_2^2}\right)+ f(x_1)\frac{(\bs{q}_{1}\cdot \hat{\bs{x}}_1)(\bs{q}_{2}\cdot \hat{\bs{x}}_1)}{q_1^2 q^2_2}(\bs{q}_{12}\cdot \hat{\bs{x}}_1)\right]\right)
\end{aligned}
\end{equation}
where $\bs{q}_{12}= \bs{q}_1+\bs{q}_2$, $b_1$  and $b_2$ are the Eulerian linear and second order clustering biases respectively, and $b_{\Gamma_2}$ is the tidal bias with the kernel,
\begin{equation}
    \mathcal{S}_2(\bs{q}_1,\bs{q}_2)= (\hat{\bs{q}}_1 \cdot \hat{\bs{q}}_2)^2 - 1
\end{equation}

As wide-separation and relativistic effects are generally relevant only on large scales, we ignore higher-order loop contributions and the phenomenological Fingers-of-God (FoG) damping that affects the bispectrum on smaller, nonlinear scales. Though we note that mixing between these contributions will still be non-negligible for the bispectrum, particularly in the squeezed limit as it correlates both large and small scales, we leave this modelling to future work.

\subsection{Relativistic effects}

Relativistic distortions occur as we observe on our past light cone. These distortions include projection type effects arising from the Doppler-type effects and gravitational redshift as well as integrated effects like the Integrated Sachs-Wolfe (ISW) and lensing contributions.

Relativistic projection effects been well studied in linear perturbation theory \cite{Yoo_2010,Bonvin_2011,Chalinor_2011}. At second order the picture is more complicated \cite{Bertacca_2014I,Bertacca_2014II,Bertacca_2015III,DiDio_2014,Yoo_2014,Fuentes_2021}, with extra couplings between the projection effects and additional dynamical contributions arising from relativistic treatment of the second order gravitational and velocity potentials. 

Here we will make use of the calculations presented in \cite{Umeh_2017,Jolicoeur_2017,Jolicoeur_2018,Clarkson_2019,deWeerd_2019}, which includes all local corrections arising from projection effects along the LOS, as well as contributions from the dynamical evolution. We neglect integrated terms in common with these works, though these will be interesting to include in a more complete analysis.

Keeping terms up to $(\mathcal{H}/k)^2$, the relativistic part of the redshift-space kernels are given by
\begin{subequations}
    \begin{align}
        Z^{(1)}_{\rm GR}=i \, &(\bs{q}_{1} \cdot \hat{\bs{x}}_1) \frac{\gamma_1}{q_1^2}+\frac{\gamma_2}{q_1^2},\\
        \begin{split}
             Z^{(2)}_{\rm GR} = \frac{1}{q^2_1 q^2_2}\bigg(& \frac{q^2_1 q^2_2}{q^2_3}\left[F_2(\bs{q}_1,\bs{q}_2) \beta_6 + G_2(\bs{q}_1,\bs{q}_2) \beta_7\right]+(\bs{q}_{1}\cdot \hat{\bs{x}}_1)(\bs{q}_{2}\cdot \hat{\bs{x}}_1)\beta_8 \\ &+ (\bs{q}_{3}\cdot \hat{\bs{x}}_1)^2 \left(\beta_9 + E_2(\bs{q}_1,\bs{q}_2)\beta_{10}\right)+ (\bs{q}_{1}\cdot \bs{q}_2)\beta_{11} + (q^2_1+q_2^2) \beta_{12} \\ &+ ((\bs{q}_{1}\cdot \hat{\bs{x}}_1)^2+(\bs{q}_{2}\cdot \hat{\bs{x}}_1)^2)\beta_{13}\\ &+ i \Big\{[(\bs{q}_{1}\cdot \hat{\bs{x}}_1)q_1^2+(\bs{q}_{2}\cdot \hat{\bs{x}}_1)q_2^2]\beta_{14}+\left[(\bs{q}_{1}\cdot \hat{\bs{x}}_1)+(\bs{q}_{2}\cdot \hat{\bs{x}}_1)\right](\bs{q}_{1}\cdot \bs{q}_2)\beta_{15}\\ & \quad+q_1 q_2[(\bs{q}_{1}\cdot \hat{\bs{x}}_1)+(\bs{q}_{2}\cdot \hat{\bs{x}}_1)]\beta_{16}+[(\bs{q}_{1}\cdot \hat{\bs{x}}_1)^3+(\bs{q}_{2}\cdot \hat{\bs{x}}_1)^3)]\beta_{17}\\ & \quad+(\bs{q}_{1}\cdot \hat{\bs{x}}_1)(\bs{q}_{2}\cdot \hat{\bs{x}}_1)[(\bs{q}_{1}+\bs{q}_{2})\cdot \hat{\bs{x}}_1]\beta_{18}+(\bs{q}_{3}\cdot \hat{\bs{x}}_1) \frac{q^2_1 q^2_2}{q^2_3}G_2(\bs{q}_1,\bs{q}_2)\beta_{19} \Big\}\bigg),
        \end{split}
    \end{align}
\end{subequations}
where
\begin{equation}
    E_2(\bs{q}_1,\bs{q}_2) = \frac{q^2_1 q^2_2}{q^4_3}\left[3+ 2\hat{\bs{q}}_1 \cdot \hat{\bs{q}}_2 \left(\frac{q_2}{q_1}+\frac{q_2}{q_2}\right)+ (\hat{\bs{q}}_1 \cdot \hat{\bs{q}}_2)^2 \right].
\end{equation}

The time dependence (e.g., dependence on radial comoving distance) is contained within the $\gamma$'s and $\beta$'s, for the first and second orders respectively, which are bias dependent coefficients evaluated at $x_i$. For example, $\gamma_1$ the leading order relativistic correction and $\gamma_2$ the second order part can be written as,
\begin{subequations}\label{eq:gamma1gamma2}
    \begin{align}
        \frac{\gamma_1(x_1)}{\mathcal{H}(x_1)} &= f(x_1)\left[b_e(x_1) - 2\mathcal{Q}(x_1)-\frac{2(1-\mathcal{Q}(x_1))}{x_1 \, \mathcal{H}(x_1)}-\frac{-(1+z(x_1))\mathcal{H}(x_1)\mathcal{H}_z(x_1)}{\mathcal{H}^2(x_1)}\right],\\
        \begin{split}
            \frac{\gamma_2(x_1)}{\mathcal{H}^2(x_1)} &= f(x_1)(3-b_e(x_1)) + \frac{3}{2}\Omega_m(x_1) \bigg[2 + b_e(x_1) - f(x_1) - 4 \mathcal{Q}(x_1)\\ & \hspace{155pt} -\frac{2(1-\mathcal{Q}(x_1))}{x_1 \, \mathcal{H}(x_1)}-\frac{-(1+z(x_1))\mathcal{H}(x_1)\mathcal{H}_z(x_1)}{\mathcal{H}(x_1)^2}\bigg].
        \end{split}
    \end{align}
\end{subequations}
Here, $\mathcal{Q}$ and $b_e$ are respectively the magnification and evolution biases and $\Omega_m$ is the matter density parameter.

The leading-order relativistic terms are projection effects arising from the redshift-space projection, with the Doppler and gravitational redshift contributions. These terms, represented by $\gamma_1$ and $\beta_{14}-\beta_{19}$, are imaginary and scale as $\mathcal{H}/k$; they are also of odd parity and, as such, they contribute to the odd multipole moments~\cite{Clarkson_2019,deWeerd_2019}. While the second-order terms that scale as $(\mathcal{H}/k)^2$ are real and contribute to the even multipoles. Full expressions for the $\beta$'s are given in Appendix A of \cite{deWeerd_2019}.

\subsection{Local tree-level bispectrum}

The redshift-space kernels, including the relativistic parts, are given by
\begin{subequations}
    \begin{align}
        Z^{(1)} &= Z^{(1)}_{\rm N} + Z^{(1)}_{\rm GR}, \\
        Z^{(2)} &= Z^{(2)}_{\rm N} + Z^{(2)}_{\rm GR}.
    \end{align}
\end{subequations}

By considering the definition of the local bispectrum (Equation~\eqref{eq:local_bispectrum}), the configuration space density fields, $\delta_s(\bs{x})$, can be expressed in terms of the inverse Fourier transform of the standard redshift-space fields, $\delta_{s} (\bs{k})$, such that the leading-order local bispectrum can be expressed as
\begin{equation}\label{eq:full_local_bk}
    \begin{aligned}
        B_{\text{loc}}(k_1,k_2,k_3,\mu_1,\phi)=\int_{\bs{x}_{13},\bs{x}_{23},\bs{q}_1,\bs{q}_2}& {\rm e}^{-i[(\bs{k}_1-\bs{q}_1) \cdot \bs{x}_{13} +(\bs{k}_2-\bs{q}_2) \cdot \bs{x}_{23}]} \\ &\times 2 Z^{(1)}(\bs{q}_1,\bs{x}_1)Z^{(1)}(\bs{q}_2,\bs{x}_2)Z^{(2)}(\bs{q}_1,\bs{q}_2,\bs{x}_3) P(q_1)P(q_2).
    \end{aligned}
\end{equation}
 
\section{Wide-Separation Corrections}\label{sec:WS_corrections}

The local bispectrum correlates the density field in three separate locations in the sky (see Equation \ref{eq:local_bispectrum}), which under the assumption of local homogeneity is modelled by a single LOS. As in the global case, this assumption is also broken locally due to redshift space distortions outside the plane parallel limit and due to the redshift evolution. Thus, describing the triplet by a singular LOS, $\bs{d}$, results in a loss of signal; in observations, the density fields at these positions are statistically different. To include the full non-linear wide-separation corrections, one would therefore need to compute a statistic dependent on all 3 end-point positions. However, just as in the power spectrum, these corrections can be approximated by introducing a series expansion to account for the deviations of the triplet positions from the chosen LOS, $\bs{d}$. 

These corrections for the position of each point in the triplet can naturally be split into angular and radial parts. The angular components are termed `wide-angle' (WA) corrections and arises as we leave the local plane-parallel limit, such that $\hat{\bs{x}}_1\neq\hat{\bs{x}}_2\neq\hat{\bs{x}}_3$; the magnitude of these corrections are therefore dependent on the size of the opening angle between the triplet and the observer, which can be defined in terms of $(x_{13}/d)$ and $(x_{23}/d)$. Corrections to the radial part arise as these statistics are measured on a light cone, and the points in the triplet are at differing comoving distances, such that $x_1\neq x_2\neq x_3$, corresponding to different redshifts. Hence, these terms depend on the size of the radial separation and on how the cosmology evolves with redshift. We introduce the term `radial redshift' (RR) to label these corrections.

Perturbative expansions to include wide-separation corrections are performed on the triangle in configuration space, and thus it is relatively straightforward to include them in the correlation functions; however, for Fourier-space statistics, the calculation is slightly more involved. The standard approach for the power spectrum is to perform the expansion in configuration space first, computing the multipoles of the two-point correlation function, and then Hankel transforming to get the multipoles of the power spectrum \cite{reimberg2016,castorina2018}. This type of approach was extended to the bispectrum in \cite{Pardede_2023}, where the angular dependencies of the three-point correlation function (3PCF) were expanded into Legendre multipoles and as such, after computing the angular integrals, the 3PCF can be related to the bispectrum multipoles by a double Hankel transform. 

In \cite{Noorikuhani_2022}, a formalism using a Cartesian space expansion is introduced. This allows for the computation of wide-separation corrections directly in Fourier space. This approach has the advantage that it connects neatly with standard perturbation theory (SPT) in Fourier space and, therefore, is easily extendable to higher-order statistics. For the bispectrum, this method also has the advantage of being fully analytic and only reliant on finite sums. Here, we use this formalism to compute wide-separation corrections, including radial-redshift corrections, for a generalised LOS. Below we describe our calculations.

\subsection{Geometry in configuration space}\label{Sec:geom_config}

\begin{figure}[tbp]
\centering 
\includegraphics[width=.75\textwidth]{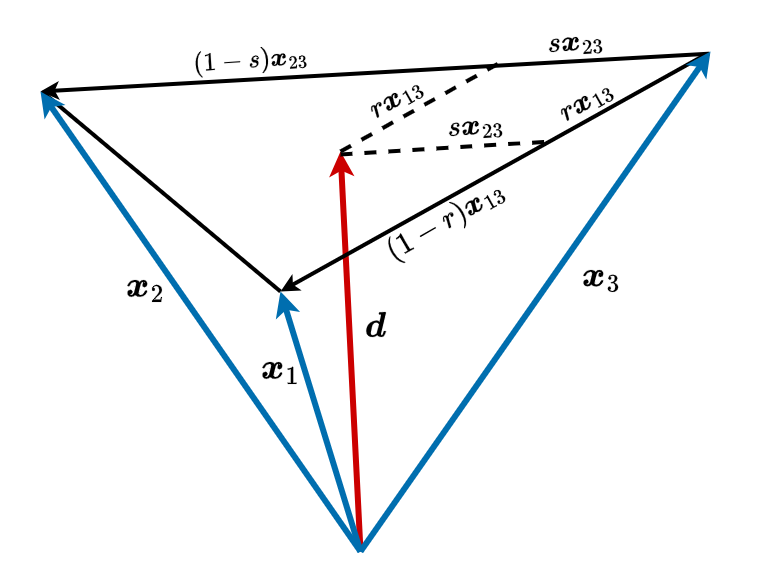}
\caption{\label{fig:real_space_bk_x3} Bispectrum geometry in real space, where the LOS $\bs{d}$ and the positions of the real space end-points LOS defined from the observer. The parameters $r$ and $s$ are the weights relating $\bs{d}$ to $\bs{x}_3$.}
\end{figure}

    The choice of LOS here then represents both the point from which we perform our expansions but also the unit vector which we use to define our spherical harmonics multipoles. This choice can be generalized to any point in the real-space triangle defined by $\bs{x}_1,\bs{x}_2,\bs{x}_3$ using two free parameters such that
    \begin{equation}\label{eq:bk_LOS}
    \bs{d} = r\,\bs{x}_1 + s\,\bs{x}_2 + (1-r-s)\,\bs{x}_3,
    \end{equation}
    where $0<r+s<1$ and are defined with respect to $\bs{x}_3$.
     
    From this parameterisation, the three end-point LOS $\bs{x}_1,\bs{x}_2$ and $\bs{x}_3$ can be re-expressed in terms of the parameters of our local bispectrum, Equation~\eqref{eq:local_bispectrum}: $\bs{d}$, at which we define our local bispectrum, and $\bs{x}_{13}$ and $\bs{x}_{23}$, which are integrated over. By inspecting Figure~\ref{fig:real_space_bk_x3}), we see that we can write, for example, \begin{equation}\label{eq:x1x2x3_expansion}
        \begin{aligned}
            \bs{x}_1 &= \bs{d} + (1-r)\bs{x}_{13} - s\bs{x}_{23},
            \\   
            \bs{x}_2 &= \bs{d} - r\bs{x}_{13} + (1-s) \bs{x}_{23},
            \\
            \bs{x}_3 &= \bs{d} - r\bs{x}_{13} - s\bs{x}_{23}.
        \end{aligned}
    \end{equation}
    Common LOS choices, such as the end-point or centre-of-mass (COM), are then defined as: 
    \begin{equation}
    \begin{aligned}
        \bs{x}_1 &: \,\, r=1,\quad s=0, \\
        \bs{x}_2 &: \,\, r=0,\quad s=1,\\
        \bs{x}_3 &: \,\, r=0,\quad s=0,\\ 
        \text{COM} &: \,\, r=s=1/3.
    \end{aligned}
    \end{equation}

    For convenience in the series expansion below, we also define the notation 
    \begin{equation}
    \begin{aligned}
    A_i &= \begin{cases}
            A_1 = (1-r), & \text{for } \bs{x}_1 \\
            A_2 = -r,    & \text{for } \bs{x}_2 \\
            A_3 = -r,    & \text{for } \bs{x}_3
           \end{cases} \\
    \end{aligned}
    \end{equation}
    and 
    \begin{equation}
    \begin{aligned}
    B_i &= \begin{cases}
            B_1 = -s,    & \text{for } \bs{x}_1 \\
            B_2 = (1-s), & \text{for } \bs{x}_2 \\
            B_3 = -s,    & \text{for } \bs{x}_3
           \end{cases}
    \end{aligned}
    \end{equation}
    where $A_i$ then represents the separation between the chosen LOS and a given end-point, $\bs{x}_i$, along the vector $\bs{x}_{13}$ and $B_i$ is the equivalent for the vector $\bs{x}_{23}$. The series expansion for an end-point dependence then can be written generally in terms of $A_i,B_i$.
    
    From these definitions, the magnitudes of each vector $x_1,x_2,x_3$ can then be found in terms of $A_i,B_i$

    \begin{equation}\label{eq:LOS_mag}
        x_i = d \sqrt{1+2 \mu_{13} A_i\epsilon_{1}+2\mu_{23}B_i\epsilon_{2}+A_i^2\epsilon^2_{1}+A_i^2\epsilon^2_{2}+2A_i B_i(\hat{\bs{x}}_{13}\cdot\hat{\bs{x}}_{23})\epsilon_{1}\epsilon_{2}},
    \end{equation}
    where $\mu_{13}= \hat{\bs{x}}_{13} \cdot \hat{\bs{d}}$ and $\mu_{23}= \hat{\bs{x}}_{23} \cdot \hat{\bs{d}}$
    and we have defined the parameters  $\epsilon_{1} = x_{13}/d \ \& \ \epsilon_{2} = x_{23}/d$. 
    
    The approach is then to Taylor expand any $\bs{x}_1,\bs{x}_2,\bs{x}_3$ dependence in terms of $\epsilon_1,\epsilon_2$ about the point $\epsilon_{1}=\epsilon_{2}=0$, to incorporate these corrections from the breaking of statistical homogeneity in our local bispectrum, \eqref{eq:local_bispectrum}. The validity of this expansion for a given survey is dependent on its range of both redshifts and scales. If we consider a full sky survey for a given redshift bin then although $\epsilon_1,\epsilon_2$ will be larger than one for some local triplets with the largest separations, the contributions of them to most k-scales we are interested in should be small. For example, if we focus on a scale of $k \approx 0.01 \, [h/ {\rm Mpc}]$ corresponding to wavelengths of $\sim 600 \, [{\rm Mpc}/h]$ ($\lambda = 2 \pi/k $) then this corresponds to the comoving distance to a triplet at $z \approx 0.2$. Therefore, if we were considering a triangle configuration with a $k$-mode $k \approx 0.01 \, [h/ {\rm Mpc}]$ then we would expect the expansion to break down if the bin included redshifts as low as  $z \approx 0.2$. So therefore to be robust, for a given redshift bin with a $z_{\rm min}$ the minimum $k$ (largest scale) that the expansion is valid for is $k_{\rm min} = 2 \pi/x(z_{\rm min})$ \cite{benabou2024}.

    \subsubsection{Wide-angle expansion}
    The angular LOS dependence in the bispectrum is included in terms of $\hat{\bs{x}}_1,\hat{\bs{x}}_2$, and $\hat{\bs{x}}_3$, which can be rewritten in terms of a dual series expansion in $\epsilon_1$ and $\epsilon_2$ about the LOS $\hat{\bs{d}}$. Using Equations \eqref{eq:x1x2x3_expansion} and \eqref{eq:LOS_mag}, the unit vectors can be expanded, up to second order in $\epsilon_1$ and $\epsilon_2$, as:
    \begin{equation}\label{eq:WA_expansion}
        \begin{split} 
          \hat{\bs{x}}_i &= \frac{\bs{d} + A_i\bs{x}_{13} + B_i\bs{x}_{23}}{|\bs{d} + A_i\bs{x}_{13} + B_i\bs{x}_{23}|}, \\ &= \left( \hat{\bs{d}} + A_i\,\hat{\bs{x}}_{13}\,\epsilon_1+B_i\,\hat{\bs{x}}_{23}\,\epsilon_2\right) 
          \bigg[1 -  A_i\,\mu_{13}\,\epsilon_1 -B_i\,\mu_{23}\,\epsilon_2- \frac{1}{2}\left(A^2_1 \, \epsilon^2_1 + A^2_2 \, \epsilon^2_2\right) \\ &\quad A_i \, B_i \, (\hat{\bs{x}}_{13} \cdot \hat{\bs{x}}_{23}) \, \epsilon_{1} \, \epsilon_{2} + \frac{3}{8} \left( 4 \, A^2_1 \, \mu^2_{13} \, \epsilon^2_1 + 4 \, A^2_2 \, \mu^2_{23} \, \epsilon^2_2 + 8 \, A_i \, B_i \, \mu_{13} \, \mu_{23} \, \epsilon_1 \, \epsilon_2 \right) \bigg] + \mathcal{O}(\epsilon^3),
          \\ &= \hat{\bs{d}} + A_i \, \hat{\bs{x}}_{13} \, \epsilon_1 + B_i \, \hat{\bs{x}}_{23} \, \epsilon_2 - \hat{\bs{d}} (A_i \, \mu_{13} \, \epsilon_1 + B_i \, \mu_{23} \, \epsilon_2) - \hat{\bs{d}} \Bigg[ \frac{1}{2} \left(A^2_1 \, \epsilon^2_1 + A^2_2 \, \epsilon^2_2 \right) \\ & \quad +A_i \, B_i (\hat{\bs{x}}_{13} \cdot \hat{\bs{x}}_{23}) \, \epsilon_{1} \, \epsilon_{2} - \frac{3}{8} \left( 4 \, A^2_1 \, \mu^2_{13} \, \epsilon^2_1 + 4 \, A^2_2 \, \mu^2_{23} \, \epsilon^2_2 + 8 \, A_i \, B_i \, \mu_{13} \, \mu_{23} \, \epsilon_1 \, \epsilon_2 \right) \Big]\\ & \qquad -(A_i \, \hat{\bs{x}}_{13} \, \epsilon_1 + B_i \, \hat{\bs{x}}_{23} \, \epsilon_2)(A_i \, \mu_{13} \, \epsilon_1 + B_i \, \mu_{23} \, \epsilon_2) + \mathcal{O}(\epsilon^3).
        \end{split}
    \end{equation}
    This can then be trivially extended to any order.
    
    \subsubsection{Radial-redshift expansion}\label{sec:rr_deriv}
    
    Radial dependence in the local bispectrum is included in the form of several redshift-dependent parameters, which are evaluated at the comoving distances ${x_1,x_2,x_3}$. Outside the constant redshift approximation, where $x_i\neq d$, it follows that $f(x_i)\neq f(d)$, where $f(x_i)$ is any term we might consider as a function of redshift, i.e., $D,f,K,C,b_1,b_2,b_{\gamma},\beta_i$ etc. If we consider Equation~\eqref{eq:LOS_mag}, the corrections to the radial part can naturally be included in the dual Taylor expansion in $\epsilon_1$ and $\epsilon_2$ about the point $\epsilon_1=\epsilon_2=0$, such that, for any function $f(x_i)$, additional terms are generated depending on its derivative with respect to radial comoving distance:
    \begin{equation}\label{eq:RR_expansion}
        \begin{aligned}
            f(x_i) = &  f(d) + A_i \, f'(d)\,\mu_{13}\,\epsilon_{1} + B_i\, f'(d)\,\mu_{23}\,\epsilon_{2} +\frac{1}{2}f'(d)\left[A_i^2 \epsilon^2_{1}+ 2 A_i\,B_i \,(\hat{\bs{x}}_{13}\cdot\hat{\bs{x}}_{23})\,\epsilon_{1}\,\epsilon_{2} + B_i^2 \epsilon^2_2 \right] \\ 
            & + \frac{1}{2}\left[-2 f'(d) +f''(d)\right]\left(A_i^2 \,\mu^2_{13}\,\epsilon^2_{1} +2 A_i\,B_i\, \mu_{13}\,\mu_{23}\,\epsilon_{1}\,\epsilon_{2} + B_i^2 \, \mu^2_{23}\,\epsilon^2_{2}\right) +\mathcal{O}(\epsilon^3).
        \end{aligned}
    \end{equation}
    We use derivatives with respect to the log-comoving distance, $'=\dd /\dd \,\ln d$, such that each order in the series expansion has the same suppression as the wide-angle terms, $1/(kd)^n$. However, alternatively we could consider $f'(d)= (\mathcal{H} d)f_z(d)$, in which case the first-order terms will have a $(\mathcal{H}/k)$ suppression. Still considering derivatives with respect to redshift rather than comoving distance, at higher orders the pure radial-redshift terms also can have a $1/(kd)$ suppression and, as such, the $n^{\rm th}$-order as $(1/(kd))^a (\mathcal{H}/k)^b$, for $a+b=n$.

\begin{figure}[tbp]
        \centering 
        \includegraphics[width=0.75\textwidth]{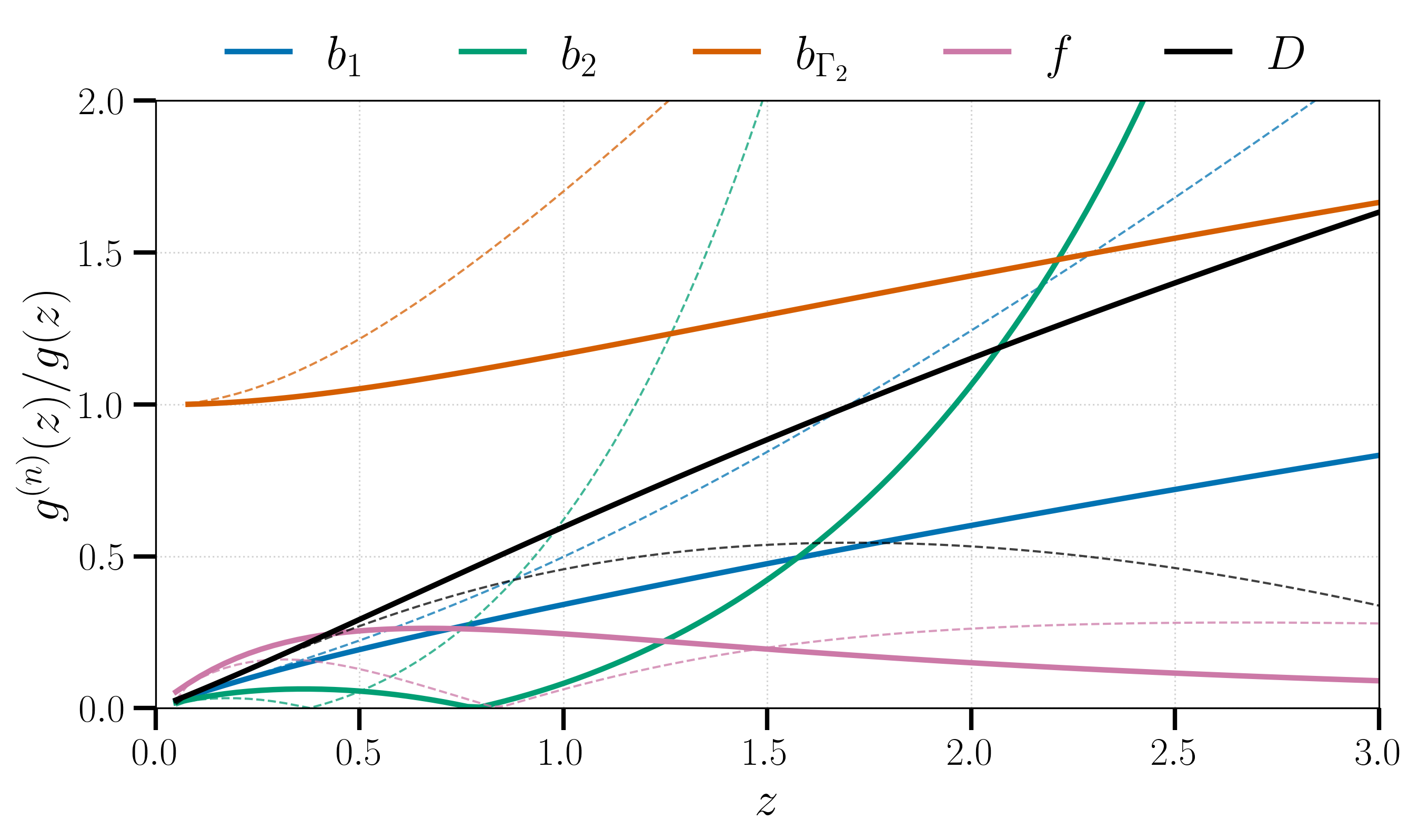}
        \caption{\label{fig:derivs} Ratios of redshift-dependent parameters with its first- (\textit{solid lines}) and second-order (\textit{dashed}) derivatives with respect to $\ln d$, as a function of redshift. In this figure, we take $b_1 = \sqrt{1+z}$ as a proxy for the linear bias, with $b_2$ and $b_{\Gamma_2}$ given by Equations~\eqref{eq:secondorderbias}. The fiducial flat $\Lambda$CDM cosmology used to compute $f$ and $D$ is described in Section \ref{sec:results}.}
    \end{figure}

    The relative size of the radial-redshift contributions to the wide-angle terms scale as $B_{\rm RR}\propto g'(d) \, B_{\rm WA}$, and thus are related to $g'(d)/g(d)$ for each parameter in the bispectrum that depends on redshift. The size of the derivative of each parameter, compared to the wide-angle terms, is shown in Figure~\ref{fig:derivs}. Note that the radial-redshift contribution from $b_{\Gamma_2}$ and $b_{2}$ is still small as these are subdominant contributions to the bispectrum at zeroth order in the wide-separation expansion.

    \subsection{Derivation Overview}

    Starting from the local bispectrum in redshift space, Equation~\eqref{eq:full_local_bk}, we can use the series expansions shown in Equations \eqref{eq:WA_expansion} and \eqref{eq:RR_expansion} to reparameterise the $\bs{x}_1,\bs{x}_2$ and $\bs{x}_3$ dependence in terms of $\bs{d}$, and $\bs{x}_{13}$ and $\bs{x}_{23}$. 
    
    Since there is no $\bs{x}_{13},\bs{x}_{23}$ dependence in the real-space integrals for the zeroth-order term in the expansion, which corresponds to the locally homogeneous limit $\bs{x}_1 =\bs{x}_2=\bs{x}_3 = \hat{\bs{d}}$, these become delta functions $\delta_{D}(\bs{q}_1-\bs{k}_1)$ and $\delta_{D}(\bs{q}_2-\bs{k}_2)$; therefore, the plane-parallel, constant redshift local bispectrum reduces to the standard expression
    \begin{equation}\label{eq:zeroorder}
        \begin{aligned}
         B^{\rm loc,0}(k_1,k_2,k_3,\mu_1,\phi) = 2 Z_1(k_1,\mu_1)Z_1(k_1,\mu_2) Z_2(k_1,k_2,\mu_3)P_{L}(k_1)P_{L}(k_2) +  \text{perms}.
        \end{aligned}
    \end{equation}

    However, beyond the zeroth order and after expanding $\bs{x}_{13}$ and $\bs{x}_{23}$ into their Cartesian vector components $x_{13,x}, \ldots, x_{23,z}$, we can collect powers of $\epsilon_{1}$ and $\epsilon_{2}$, which leads to
    \begin{equation}\label{eq:bk_int_expansion}
    \begin{split}
        B^{\text{loc}}(k_1,k_2,k_3,\mu_1,\phi,d) = &\int_{\bs{q}_1,\bs{q}_2,\bs{x}_{13},\bs{x}_{23}}  e^{-i(\bs{k}_1-\bs{q}_1) \cdot \bs{x}_{13}} e^{-i(\bs{k}_2-\bs{q}_2) \cdot \bs{x}_{23}} \\ &\quad\times\sum_{i_x,i_y,i_z}\epsilon_1^{(i_x+i_y+i_z)}\left[\left(\frac{x_{13,x}}{x_{13}}\right)^{i_x}\left(\frac{x_{13,y}}{x_{13}}\right)^{i_y}\left(\frac{x_{13,z}}{x_{13}}\right)^{i_z} \right]\\ &\quad\times \sum_{j_x,j_y,j_z}\epsilon_2^{(i_x+i_y+i_z)}\left[\left(\frac{x_{23,x}}{x_{23}}\right)^{i_x}\left(\frac{x_{23,y}}{x_{23}}\right)^{i_y}\left(\frac{x_{23,z}}{x_{23}}\right)^{i_z} \right]\\&\quad\times \mathcal{C}_{i_x,i_y,i_z,j_x,j_y,j_z}(\bs{q}_1,\bs{q}_2,\bs{q}_3,\bs{d}),
    \end{split}
    \end{equation}
    where $\mathcal{C}_{i_x,i_y,i_z,j_x,j_y,j_z}(k_1,k_2,k_3,\mu_1,\phi)$ are the bispectrum coefficients for each term in the expansion. Here, $i_x,i_y,i_z,j_x,j_y,j_z$ correspond to the power of each respective Cartesian component of $\bs{x}_{13} \, \& \, \bs{x}_{23}$, and $n = i_x+i_y+i_z+j_x+j_y+j_z$ is the order of the expansion. In the $n=0$ case, $\mathcal{C}_{0,0,0,0,0,0}$ is just the standard zeroth-order bispectrum as in Equation~\eqref{eq:zeroorder}: 
    \begin{equation}                        
        \mathcal{C}_{0,0,0,0,0,0}\left(\bs{k}_1,\bs{k}_2,\bs{k}_3,\bs{d}\right)=2Z_1(\bs{k}_1,\bs{d})Z_1(\bs{k}_2,\bs{d})Z_2(\bs{k}_1,\bs{k}_2,\bs{d})P_{L}(k_1)P_{L}(k_2) +  \text{perms}. 
    \end{equation}
    
    For $n>0$, one can remove the $\bs{x}_{13},\bs{x}_{23}$ dependence in the real-space integrals by considering the Fourier relation
    \begin{equation}
        \frac{\partial}{\partial k_j}F(\bs{k}) = -i \int \dd^3 \bs{x}\,\, e^{-i \bs{k}\cdot\bs{x}} x_j f(\bs{x}),
    \end{equation}
    such that the Cartesian components of $\bs{x}_{13} \, ,\, \bs{x}_{23}$ can be replaced with  their derivatives, i.e. $x_{13,j} \rightarrow -i\partial_{k_{1,j}}$ and $x_{23,j} \rightarrow -i\partial_{k_{2,j}}$, acting on the whole expression. Therefore, the $\bs{x}_{13}$ and $\bs{x}_{23}$ integrals, as in the locally homogeneous limit, become
    Dirac-deltas $\delta_{D}(\bs{q}_1-\bs{k}_1)$ and $\delta_{D}(\bs{q}_2-\bs{k}_2)$. 
    
    Therefore, Equation~\eqref{eq:bk_int_expansion} can be rewritten in the form 
    \begin{equation}\label{eq:WSfullexpansion}
    \begin{split}
        B^{\rm loc}(k_1,k_2,k_3,\mu_1,\phi;d) = \sum_n \left(\frac{i}{d}\right)^{n} \, \, \, \sum^{i_x+i_y+i_z+j_x+j_y+j_z=n}_{i_x,i_y,i_z,j_x,j_y,j_z}&\left(\partial^{i_x}_{k_{1,x}}\partial^{i_y}_{k_{1,y}}\partial^{i_z}_{k_{1,z}} \right)\left(\partial^{j_x}_{k_{2,x}}\partial^{j_y}_{k_{2,y}}\partial^{j_z}_{k_{2,z}} \right)\\ &\times \mathcal{C}_{i_x,i_y,i_z,j_x,j_y,j_z}\left(\bs{k}_1,\bs{k}_2,\bs{k}_3,\bs{d}\right)
        \end{split}
    \end{equation}
    such that there are $n$ partial derivatives acting on each coefficient for each term in the series. 

    Using Equation \eqref{eq:WSfullexpansion}, we can collect terms at each expansion order, which are suppressed by a factor of $i/(k\,d)$. Therefore, if we separate wide-angle and radial-redshift terms as well as the Newtonian and relativistic contributions, we can express the bispectrum as a series of terms some suppression factors,
    \begin{equation}\label{eq:fullterms}
        \begin{aligned}
        B^{\rm loc} =& \, B_N +\left(\frac{i}{k \, d}\right) (B_{{\rm WA}_1}+B_{{\rm RR}_1})+\frac{i\,{\cal H}}{k}B_{{\rm GR}_1} + \left(\frac{i}{k \, d}\right)^2 (B_{{\rm WA}_2}+B_{{\rm RR}_2}+B_{{\rm WA}_1 {\rm RR}_1})\\ & + \frac{i\,{\cal H}}{k}\left(\frac{i}{k \, d}\right)(B_{{\rm WA}_1 {\rm GR}_1}+B_{{\rm RR}_1 {\rm GR}_1}) + \left(\frac{i\,{\cal H}}{k}\right)^2 B_{{\rm GR}_2}
        \end{aligned}
    \end{equation}
    where we have truncated terms that are suppressed by ${\cal O}(1/k^3)$ and higher. The theoretical multipoles induced by wide separations, for a given survey, can then be retrieved from the full local expression, as in Equation~\eqref{eq:bk_multipole}. Here, the imaginary first-order terms have odd powers of $\mu$ and therefore enter into the odd multipoles, while the real second-order terms are of even parity.

    \section{Results and Discussion}\label{sec:results}

    Our main results are the complete expressions for the local tree-level bispectrum (Equation~\eqref{eq:fullterms}), including wide-angle, radial-redshift, and relativistic projection effects up to second order for a generalised LOS $\bs{d}$ (Section \ref{sec:WS_corrections}). For a typical analysis, this is decomposed into LOS multipoles and averaged over the survey area (see Section~\ref{sec:local}). Because the full analytic expressions are extremely long and cumbersome, we do not show them here. Instead, we plot multipoles of the bispectrum for some given shapes and scales, and introduce \textsc{CosmoWAP} \href{https://github.com/craddis1/CosmoWAP}{\faGithub} a \textit{Python} package that implements these results as well as providing the corresponding \textit{Mathematica} \href{https://github.com/craddis1/MathWAP}{\faGithub} routines used to compute them analytically. \footnote{Similar materials for the wide-separation and relativistic power spectrum multipoles (including for the multi-tracer case) are also included for completeness in both repositories.}
    
    At first order, our numerical results for the pure wide-angle bispectrum is in agreement with both \cite{Noorikuhani_2022} and \cite{Pardede_2023}; however, at second order, we only find agreement with \cite{Pardede_2023} when the comparison was possible\footnote{At second order our results for the wide angle contribution is of the same order of magnitude and has similar shape dependence to the plots in Figure 10 of \cite{Noorikuhani_2022} but we do find exact agreement as we do with the results of \cite{Pardede_2022}}.

    To examine these effects, we consider a few different types of future galaxy surveys of different tracers and redshift ranges, a  H$\alpha$ Euclid-like survey $(0.9<z<1.8)$, a DESI-like BGS $(0.05<z<0.6)$ and SKAO1 HI galaxy survey $(0<z<0.5)$ \cite{SKA} as well as a futuristic Phase-2 survey, SKAO2, galaxy survey over $(0.1<z<2)$. We show the results for a flat $\Lambda$CDM cosmology with parameters taken from \cite{Planck_2018}:  $h=0.6766,\,\Omega_{\rm b} h^2=0.02242,\,\Omega_{\rm c} h^2=0.11933,\,A_s=2.105\times10^{-9},\,n_s=0.9665$.
    
    We use these surveys as \cite{Maartens_2022} provides readily available models of magnification, $\mathcal{Q}$, and evolution, $b_e$, bias which are important for the relativistic contributions. These are astrophysical parameters that are defined with respect to the luminosity cut for a flux limited survey,
    \begin{equation}
        b_e = -\frac{\partial \, {\rm ln} \,n_g}{\partial \, {\rm ln}(1+z)}, \, \, \mathcal{Q}= - \frac{\partial \, {\rm ln} \, n_g}{\partial \, {\rm ln} \, L}\bigg|_c,
    \end{equation}
    where $L$ is the luminosity, $n_g$ is the comoving number density and $|_c$ refers to an evaluation at the flux cut. Fitting functions of evolution and magnification bias (as well as the number density) are derived from an assumed luminosity function for a given tracer and survey (see \cite{Maartens_2022} for full details).

    We assume that the linear clustering bias is independent of luminosity \cite{Maartens_2020}
    \begin{equation}
        \frac{\partial b_1}{\partial \, {\rm ln} \, L}\Big|_c =0.
    \end{equation}
    For an H$\alpha$ Euclid-like survey, we assume a 15,000 $\rm{deg}^2$ footprint with a linear bias \cite{Maartens_2020}, 
    \begin{equation}
        b_1(z) = 0.9 + 0.4 z,
    \end{equation}
    and for magnification $\mathcal{Q}$ and evolution $b_e$ biases we adopt Model 3 from \cite{Maartens_2022}, such that,
    \begin{subequations}
        \begin{align}
            \mathcal{Q}(z) &= 0.583 + 2.02\, z - 0.568\, z^2 +0.0411\, z^3, \\ 
            b_{e}(z) &= -7.29 +0.470\, z+1.17\, z^2 -0.0290\, z^3,
        \end{align}
    \end{subequations}  
    which is valid over the redshift range $(0.9<z<1.8)$. 
    
    For the 15,000 $\rm{deg}^2$ Bright Galaxy Sample (BGS) with DESI like specifications, we assume a linear bias \cite{desi} 
    \begin{equation}
        b_1(z) = 1.34/D(z),
    \end{equation}
    with magnification $\mathcal{Q}$ and evolution $b_e$ biases \cite{Maartens_2022},
    \begin{subequations}
        \begin{align}
            \mathcal{Q}(z) &= 0.282 + 2.36\, z + 2.27\, z^2 + 11.1\, z^3, \\ 
            b_{e}(z) &= -2.25 - 4.02\, z+0.318\, z^2 - 14.6\, z^3,
        \end{align}
    \end{subequations}
    valid over the redshift range $(0.05<z<0.6)$.

    For the 5,000 $\rm{deg}^2$ SKAO1 HI galaxy survey, we use a linear bias model from \cite{Yahya_2015,Bull_2016}
    \begin{equation}
        b_1(z) = 0.616 \, \rm{e}^{1.0117 \, z},
    \end{equation} 
    and similarly for the 30,000 $\rm{deg}^2$ SKAO2:
    \begin{equation}
        b_1(z) = 0.554 \, \rm{e}^{0.783 \, z}.
    \end{equation}
    Expressions for evolution and magnification biases as well as the number densities are interpolated from Table 1 and Table 2 in \cite{Maartens_2022}.

    Then for simplicity in our modelling we assume the local Lagrangian expression for $b_{\gamma_2}$ and the quadratic fit for $b_2$ \cite{Lazeyras_2016}:
    \begin{subequations}\label{eq:secondorderbias}
        \begin{align}
           b_{\Gamma_2} &= - \frac{2}{7}(b_1(z)-1), \\
           b_2 &= 0.412-2.143\, b_1(z) +0.929\, b_1^2(z) +0.008\, b_1^3(z) + 4/3 \, b_{\Gamma_2}(z).
        \end{align}   
    \end{subequations}
    
    We plot the bispectrum for differing choices of LOS. For the results of the monopole we use a COM LOS ($r=s=1/3$) as this allows for a more accurate representation of wide-separation effects (see discussion in Section~\ref{sec:los_dependence}) while for the other multipoles, where unspecified, we use $\bs{d}=\bs{x}_3$ ($r=s=0$) as a default choice as an end-point LOS has a more simple connection to the standard FFT estimators used to measure the multipoles of the bispectrum. The $\bs{x}_2$ and $\bs{x}_3$ end-points are equivalent if the respective $k$-vectors also switch, i.e. $B_{\ell}(k_1,k_2,k_3;\bs{x}_2)\equiv B_{\ell}(k_1,k_3,k_2;\bs{x}_3)$; but this symmetry is broken for the $\bs{x}_1$ case for $\ell>0$ due to the fact that the multipole expansion is about $\bs{k}_1$.

    For plots over fixed triangle shapes and scales, we choose three different triangle configurations:
    \begin{itemize}
        \item Equilateral: $k_1=k_2=k_3=0.01\: [h/ {\rm Mpc}]$
        \item Folded: $k_1=2 \, k_2=2\, k_3=0.02\: [h/ {\rm Mpc}]$
        \item Squeezed: $k_1= k_2= 10\, k_3=0.05\: [h/ {\rm Mpc}]$
    \end{itemize}
    which are used in the plots below.
 
    \subsection{Line-of-sight dependence}\label{sec:los_dependence}

    \paragraph{Monopole}

    The monopole of the bispectrum corresponds to the case where we have averaged over the LOS dependence, $\mu_1,\phi$. Therefore, it is not decomposed with respect to any LOS and as such there is no LOS dependence in our estimator; in our theory however, we still have to define a point from which to Taylor expand about to describe our configuration space triangle and, therefore, our LOS dependent describe the choices of points about which to do the expansion.

    \begin{figure}[tbp]
        \centering 
        \includegraphics[width=\textwidth]{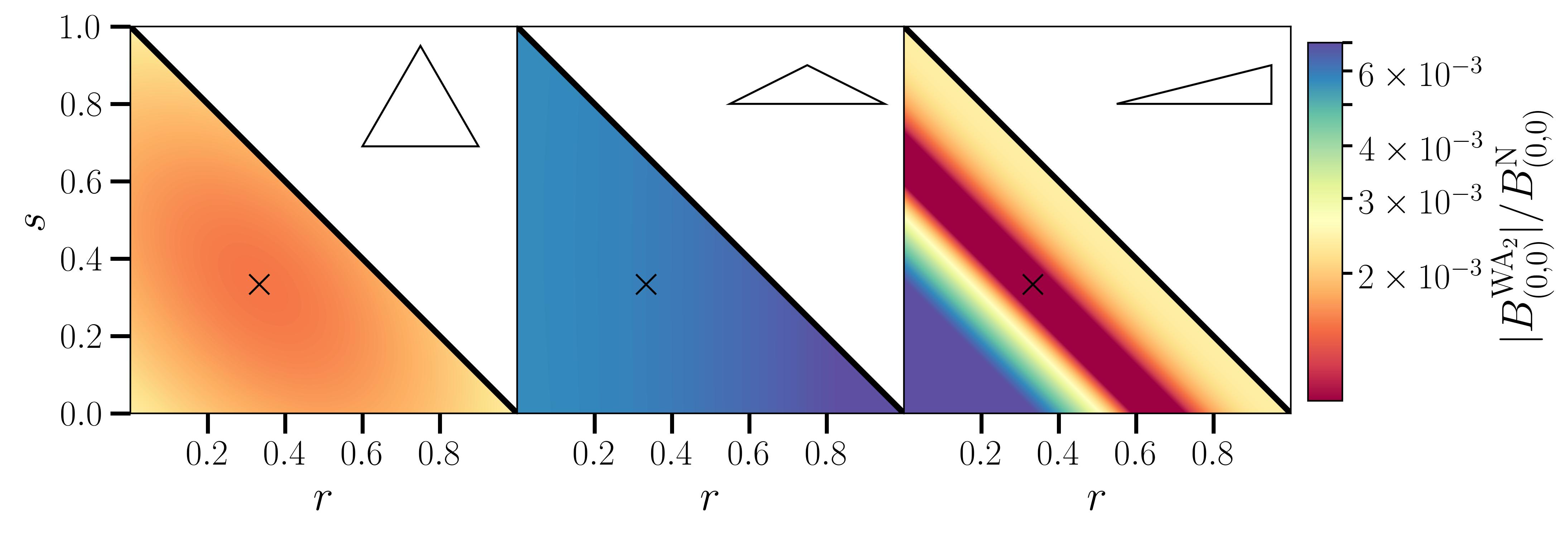}\caption{\label{fig:rs_monopole} Fractional contribution to the monopole from second-order wide-angle effects, $B^{{\rm WA}_2}_{(0,0)}$, for a H$\alpha$ Euclid-like survey at redshift $z=1$. This is plotted over different LOS directions in the configuration space triangle (see Figure~\ref{fig:real_space_bk_x3}), for three fixed triangles: equilateral (\textit{left}), folded (\textit{middle}) and squeezed (\textit{right}). The corners of the triangles represent the end-point lines-of-sight: $\bs{x}_1$ (\textit{bottom right}), $\bs{x}_2$ (\textit{top left}), and $\bs{x}_3$ \textit{(bottom left}). The COM LOS is denoted by $\times$.}
    \end{figure}
       
    For the monopole only the even order wide-separation and relativistic terms enter and so in this case we are referring purely to the second-order terms. Figure \ref{fig:rs_monopole} shows the dependence of the monopole on $r,s$ (which parameterizes our choice of LOS) for different bispectrum shapes. For the equilateral shape, $k_1=k_2=k_3$, all three end-points are equivalent, however for a non-endpoint LOS, for example, the centre-of-mass (COM) where $r=s=1/3$, will generally induce a smaller contribution. This can be explained if we consider the configuration space triangle, Figure~\ref{fig:real_space_bk_x3}; the separations between $\bs{d}$ and the end-points are less extreme for a point in the centre of the triangle. Thus, the wide-separation series expansions in terms of $A_i \epsilon_1, B_i \epsilon_2$ should converge quicker, and therefore a COM LOS should be more accurate to the true non-linear wide-separation corrections that we observe. Therefore, for all discussion of the monopole, we use a COM LOS.

    Outside the equilateral configuration, wide-separation effects are often weighted more if the LOS choice is weighted more towards an end-point which corresponds to a smaller $k$-vector; wide-separation effects are larger when the scales that are correlated are larger. 
    
    \begin{figure}[tbp]
    \centering 
    \begin{subfigure}{\textwidth}
        \centering
        \includegraphics[width=\linewidth]{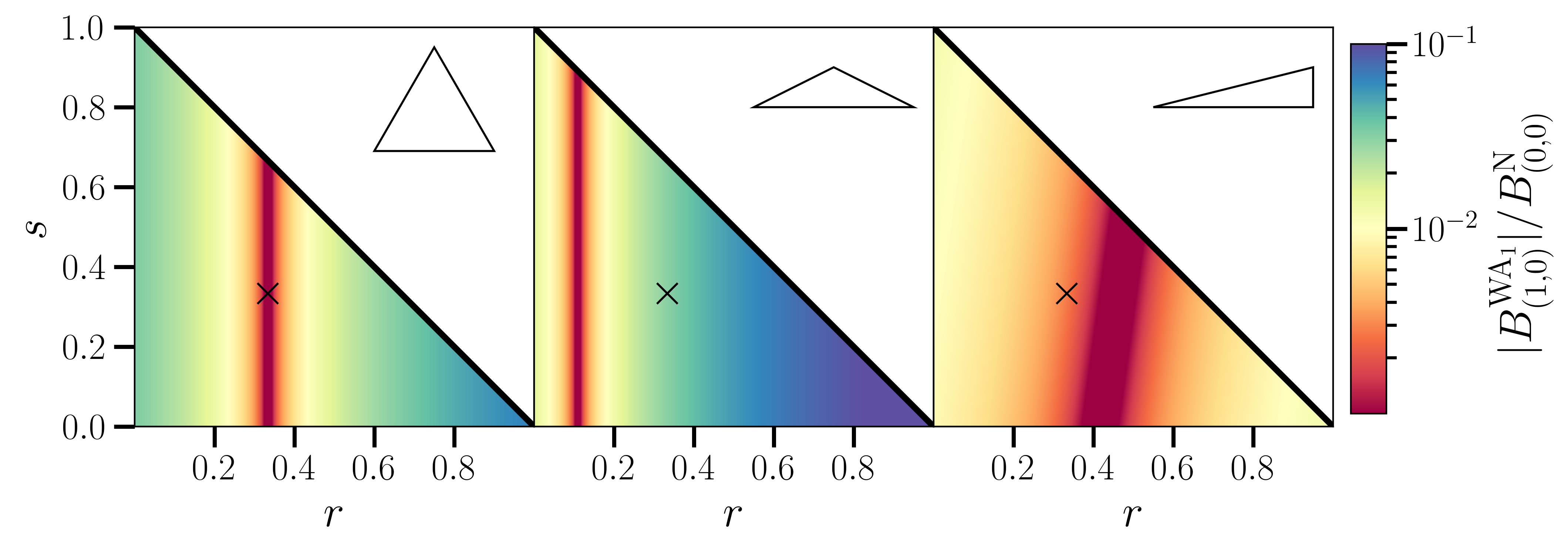}
        \caption{Wide-angle (WA) contribution.}

    \end{subfigure}
    
    \vspace{0.1em} 
    
    \begin{subfigure}{\textwidth}
        \centering
        \includegraphics[width=\linewidth]{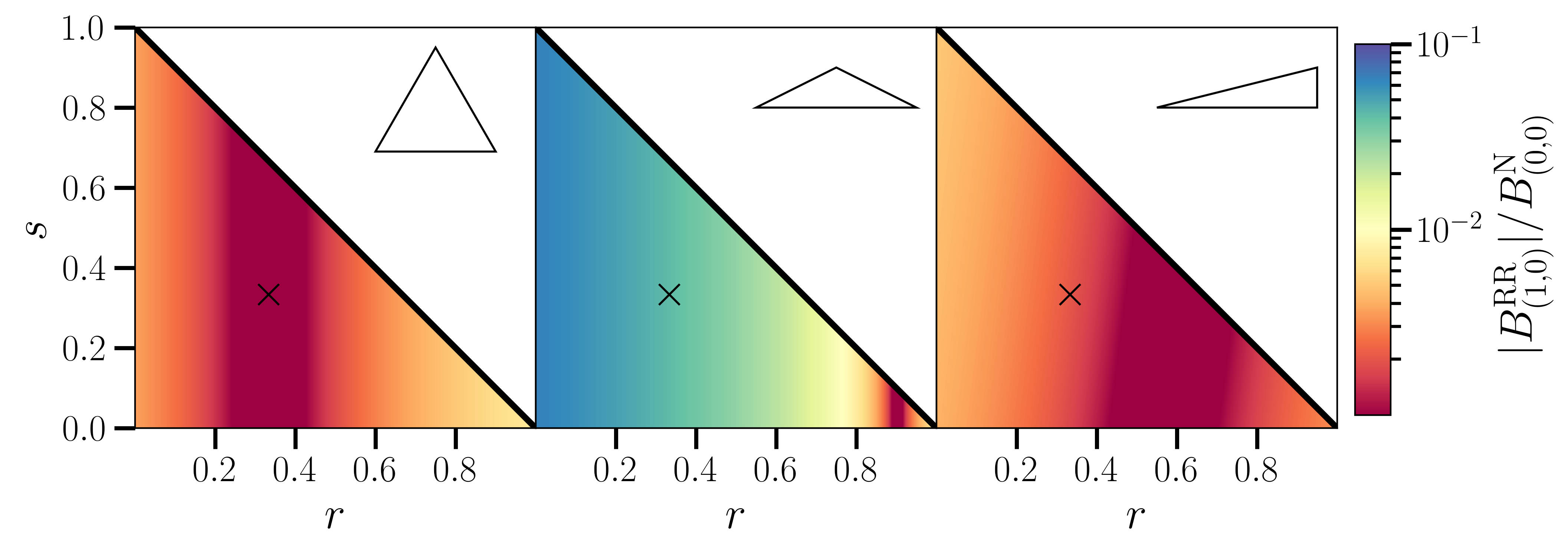}
        \caption{Radial-redshift (RR) contribution.}
    \end{subfigure}
    \caption{\label{fig:rs_dipole} Contribution to the dipole as a fraction of the monopole from first-order wide-separation effects for a Euclid-like H$\alpha$ survey at $z=1$. This is plotted over LOS choices (parameterised by $r$ and $s$, as in Equation~\ref{eq:x1x2x3_expansion}) in configuration space, for three fixed triangles: equilateral (\textit{left}), folded (\textit{middle}) and squeezed (\textit{right}). The corners of the triangles represent the end-point LOS: $\bs{x}_1$ (\textit{bottom right}), $\bs{x}_2$ (\textit{top left}), and $\bs{x}_1$ (\textit{bottom left}). The COM LOS is denoted by $\times$.}
   \end{figure}

   \paragraph{Other multipoles} For any $\ell \neq 0$ multipole, the choice of LOS is more complex as is it not only the point about which we expand from to describe the configuration space triangle, but also it is the vector with which we use to define our spherical harmonic basis. Figure~\ref{fig:rs_dipole} shows the odd wide-separation contributions (both radial evolution and wide-angle) to the bispectrum dipole as a function of $r,s$. First order wide-separation terms are linear in the expansions parameters and so there is no $r,s$ dependence from the convergence in the equilateral limit, but as before different $k$ generate $r,s$ dependence due to the loss of symmetry (for this folded configuration $k_2=k_3$ and therefore there is no change with respect to $s$). But outside the monopole there is a dependence on $r$ as our spherical harmonics are defined with respect to $\hat{\bs{k}}_1$. For the case of non-zero $m$ multipoles these also contain information about $k_2$ and therefore these have additional $s$ dependence.

    \subsection{Parity odd}

\begin{figure}[tbp]
        \centering 
        \includegraphics[width=\textwidth]{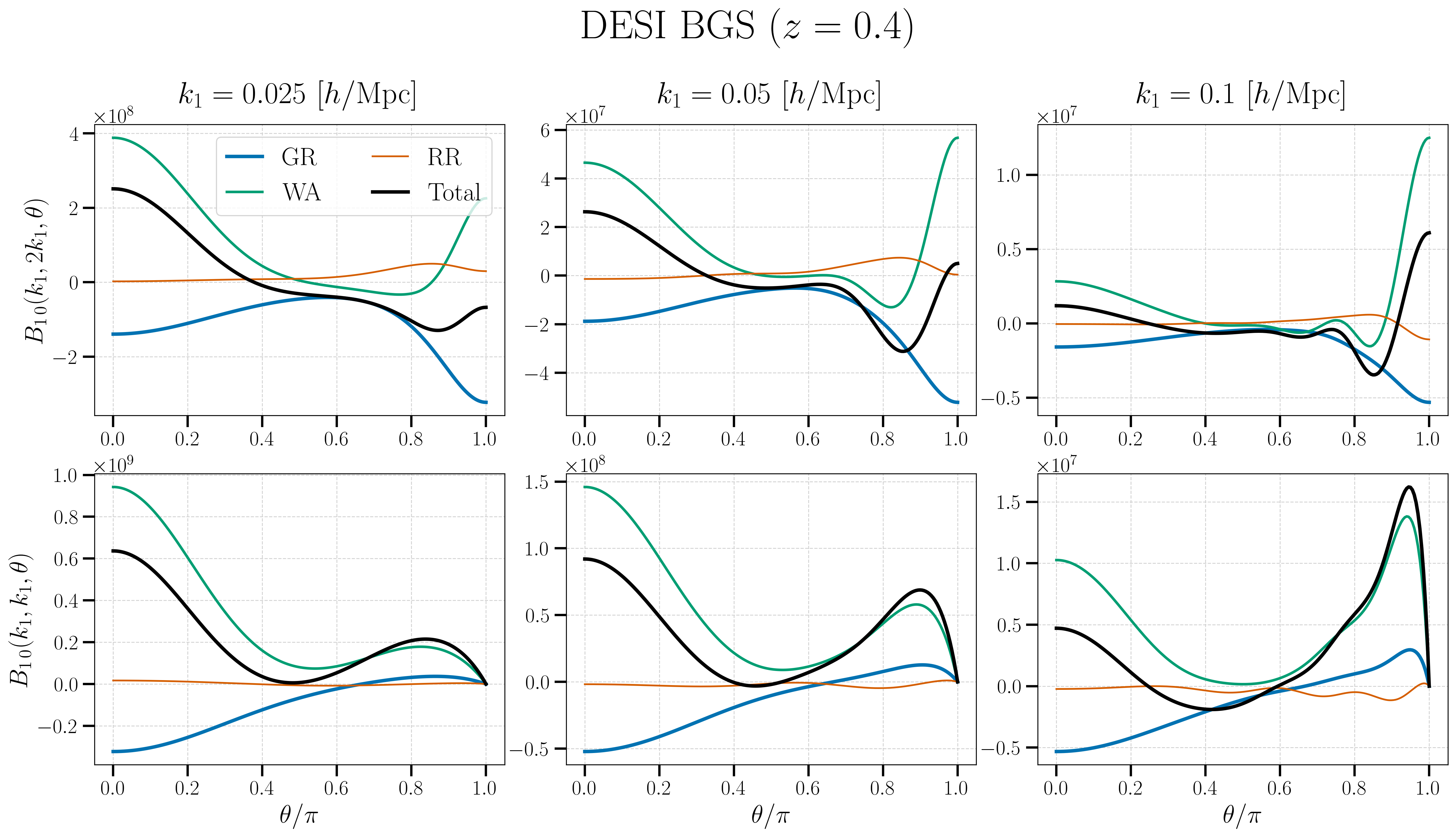}\caption{\label{fig:full_dipole_z04} Dipole moment $\ell = 1$ ($m = 0$) of the bispectrum for $k_2 = 2k_1$ (\textit{top}) and the isosceles configuration, $k_1 = k_2$ (\textit{bottom}), as a function of the angle between $k_1$ and $k_2$. The results assume a DESI-like Bright Galaxy Survey (BGS) at redshift $z = 0.4$. Each panel displays the general relativistic (GR, \textit{blue}), wide-angle (WA, \textit{green}), and radial-redshift (RR, \textit{orange}) contributions to the bispectrum, and the total signal (\textit{black}). The total bispectrum is computed from the sum of each of these contributions.}
    \end{figure}

    The imaginary contributions coming from odd moments is composed of the first-order contributions from the three different effects: wide-angle, radial-redshift and relativistic terms:
    \begin{equation}
        B^{(1)}_{\rm loc}= B_{\rm WA_1}+B_{\rm RR_1}+B_{\rm GR_1}.
    \end{equation}
    These will enter the odd multipole moments of the bispectrum, namely $B_{10}$. We show each of these contributions, and the total bispectrum, in Figure~\ref{fig:full_dipole_z04}, after integrating over all possible LOS.
 
    At first order in the wide-separation series expansion, the contributions are linear in the expansion parameters and, therefore, any choice of LOS in the configuration-space triangle can be expressed as the linear combination of the three end-point lines-of-sight:
    \begin{equation}
    \begin{aligned}
        B^{(1)}(k_1,k_2,k_3,\mu_1,\phi,d;\bs{d}) = \, &rB^{(1)}(k_1,k_2,k_3,\mu_1,\phi,d;\bs{x}_1)+ sB^{(1)}(k_1,k_2,k_3,\mu_1,\phi,d;\bs{x}_2)\\ &+ (1-r-s)B^{(1)}(k_1,k_2,k_3,\mu_1,\phi,d;\bs{x}_3).
    \end{aligned}
    \end{equation}
    
    \begin{figure}[tbp]
    \centering 
    \includegraphics[width=\textwidth]{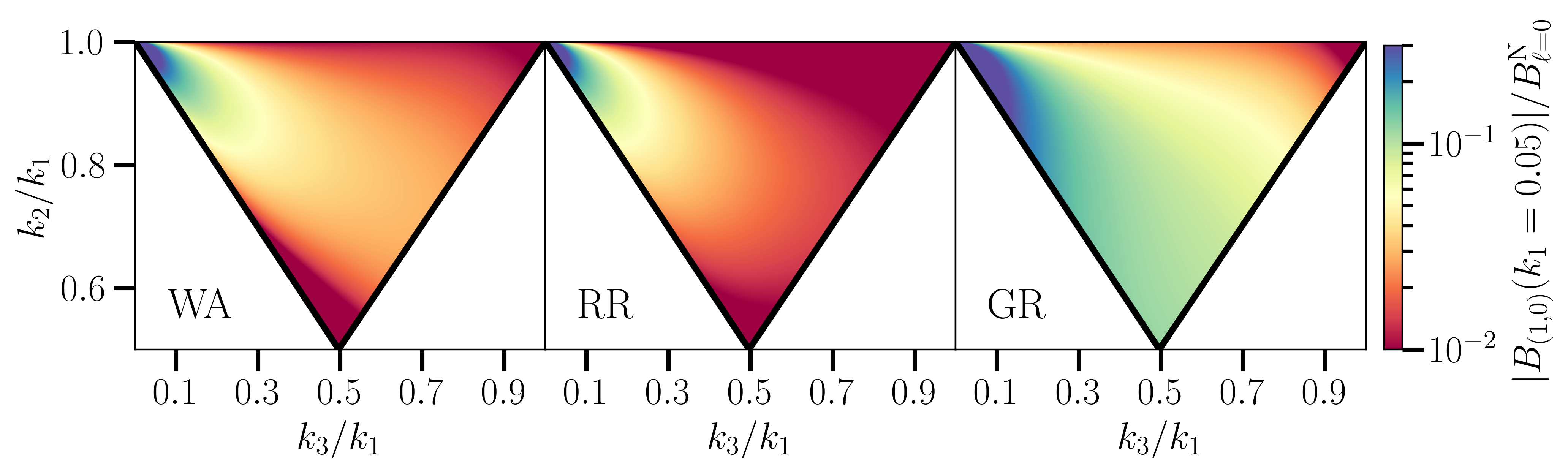}\caption{\label{fig:first_triangles} Contribution to the dipole as a fraction of the monopole from first order wide-angle, radial-redshift and relativistic effects for a H$\alpha$ Euclid-like survey at $z=1$. This is plotted over bispectrum shapes, for a fixed scale with $k_1=0.05 \, \, [h/$Mpc]. Top left corner: Squeezed, Top right corner: Equilateral, Bottom: Folded. }
    \end{figure}

    \begin{figure}[tbp]
    \centering 
    \includegraphics[width=\textwidth]{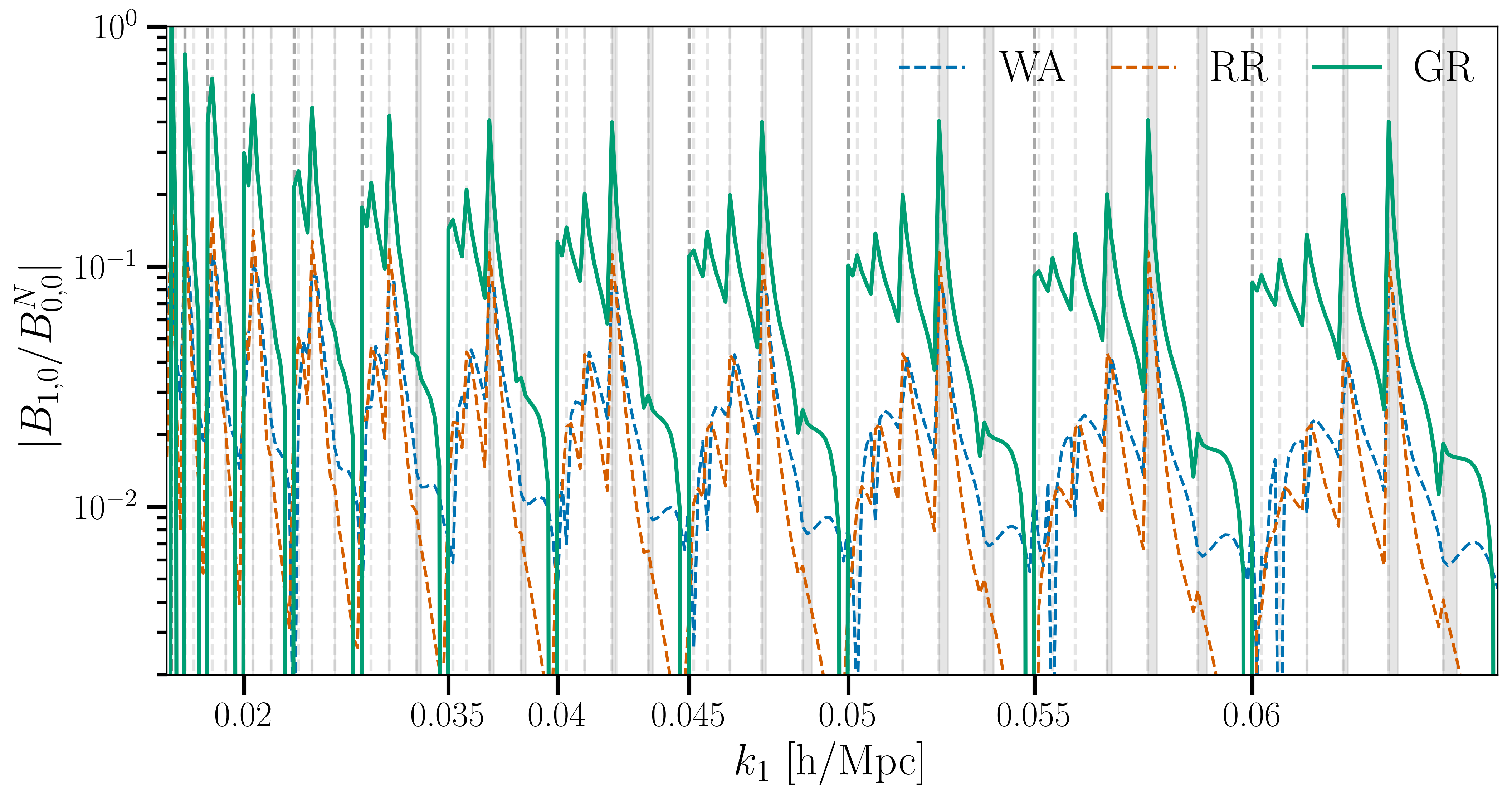}
    \caption{\label{fig:dipole_all} Odd parity first order contributions to the dipole as a fraction of the monopole plotted over all triangles, where $k_1>k_2>k_3$, for a Euclid-like H$\alpha$ survey at $z=1$. The plot represents all triangles in a nested structure where we use a bin width of $0.005\: [h/ {\rm Mpc}]$. The thicker dashed lines represent a step in $k_1$ ($k_2$ (and $k_3$) reset to their minimum value for that $k_1$ considering the triangle condition), such that it is constant in that region between dashed lines. Folded triangles occur on these dashed lines, with equilateral triangles directly before it. The fainter dashed lines denote the corresponding steps in $k_2$. The shaded regions represent `squeezed' regions where $3 \, k_3 < k_2 \leq k_1$.}
    \end{figure}

    As expected, due to their $1/k$ suppression, Figure~\ref{fig:first_triangles} and Figure~\ref{fig:dipole_all} shows that the signal for all three effects increases at large scales with peaks in the squeezed limit, corresponding to a small $k_3$ in this case.  A notable result here is that relativistic effects are in general larger than their wide-separation counterparts (note that for a single tracer in the equilateral case, the relativistic effects are zero due to the symmetry), unlike in the power spectrum, though we stress the relativistic terms are heavily dependent on the models of evolution and magnification bias. Wide-separation effects can also lead to higher powers of $\mu$ and, as such, it has the consequence of effectively moving the signal into higher-order multipoles and therefore for higher $\ell$, wide-separation effects will generally have a greater fractional contribution to the overall signal. We include plots of the contributions for other odd multipoles in Appendix~\ref{ap:other_multipoles}.
    
\begin{figure}[tbp]
    \centering 
    \includegraphics[width=\textwidth]{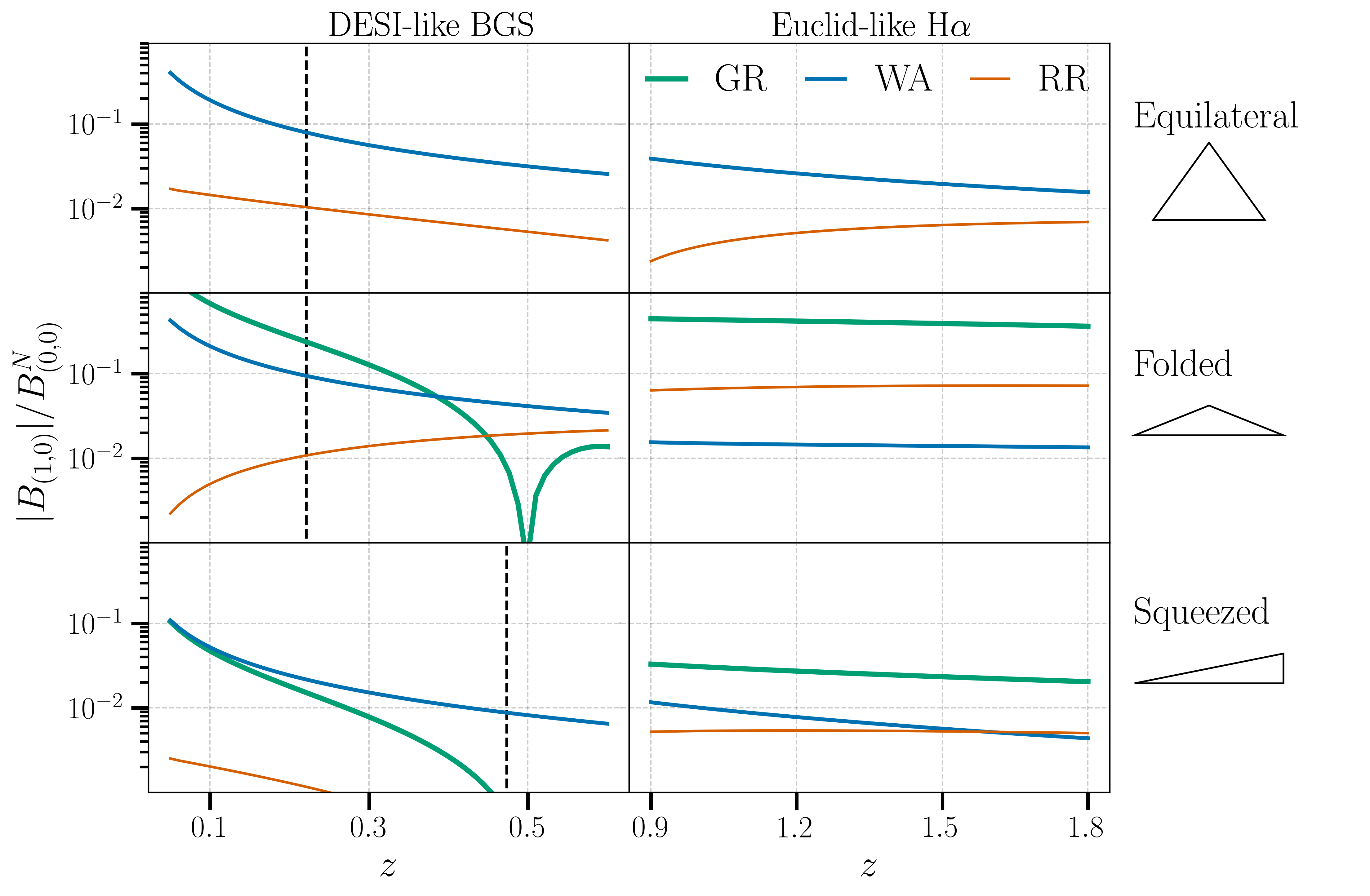}
    \caption{\label{fig:redshift_first} Odd parity first-order contributions to the dipole as a fraction of the monopole for three different triangle configurations (equilateral, squeezed and folded) for both a DESI BGS and an H$\alpha$ Euclid-like survey. The dashed vertical represent the redshift at which the wide-separation expansion is expected to break down for the given scales.}
    \end{figure}

    The wide angle contribution has a $1/(kd)$ suppression and therefore it is less important at high redshifts, as shown in Figure~\ref{fig:redshift_first}. However, for the BGS, at low redshifts, they can dominate the dipole signal, though at a certain redshift and scale, given by $k = 2 \pi/\chi(z)$, the wide-separation perturbative expansion will break down (represented by the dashed vertical lines in Figure~\ref{fig:redshift_first}). 
    
    First-order radial-redshift contributions however, if we use derivatives with respect to redshift, scale as $\mathcal{H}/k$. Therefore, for surveys covering a higher redshift range, such as the Roman Space Telescope \cite{wfirst}, or the proposed Stage-V MegaMapper survey \cite{Megamapper}, we would expect the wide-angle contribution to be negligible, while the radial-redshift corrections would still need to be considered for precision analyses.

    \subsubsection{Signal-to-noise ratio}

    To consider the detectability of these effects in future surveys we compute the signal-to-noise ratio (SNR) for the dipole of the bispectrum:
    \begin{equation}
        {\rm SNR}^2=\sum_{\triangle} \frac{B_{\ell m}(k_1,k_2,k_3)B_{\ell m}^*(k_1,k_2,k_3)}{{\sf C}_{\ell m}(k_1,k_2,k_3)}
    \end{equation}
    where we discretely sum over all triangles $\triangle$ where $k_1 > k_2 > k_3$\footnote{This is imposed to only count unique triangles, though wide-separation effects partially break the symmetry; for example, the results would be identical if we imposed $k_1 > k_3 > k_2$ and used an $\bs{x}_2$ as our LOS choice.}. We assume a Gaussian bispectrum covariance for simplicity with the expressions given in Appendix \ref{ap:covariance}.

\begin{figure}[tbp]
    \centering 
    \includegraphics[width=\textwidth]{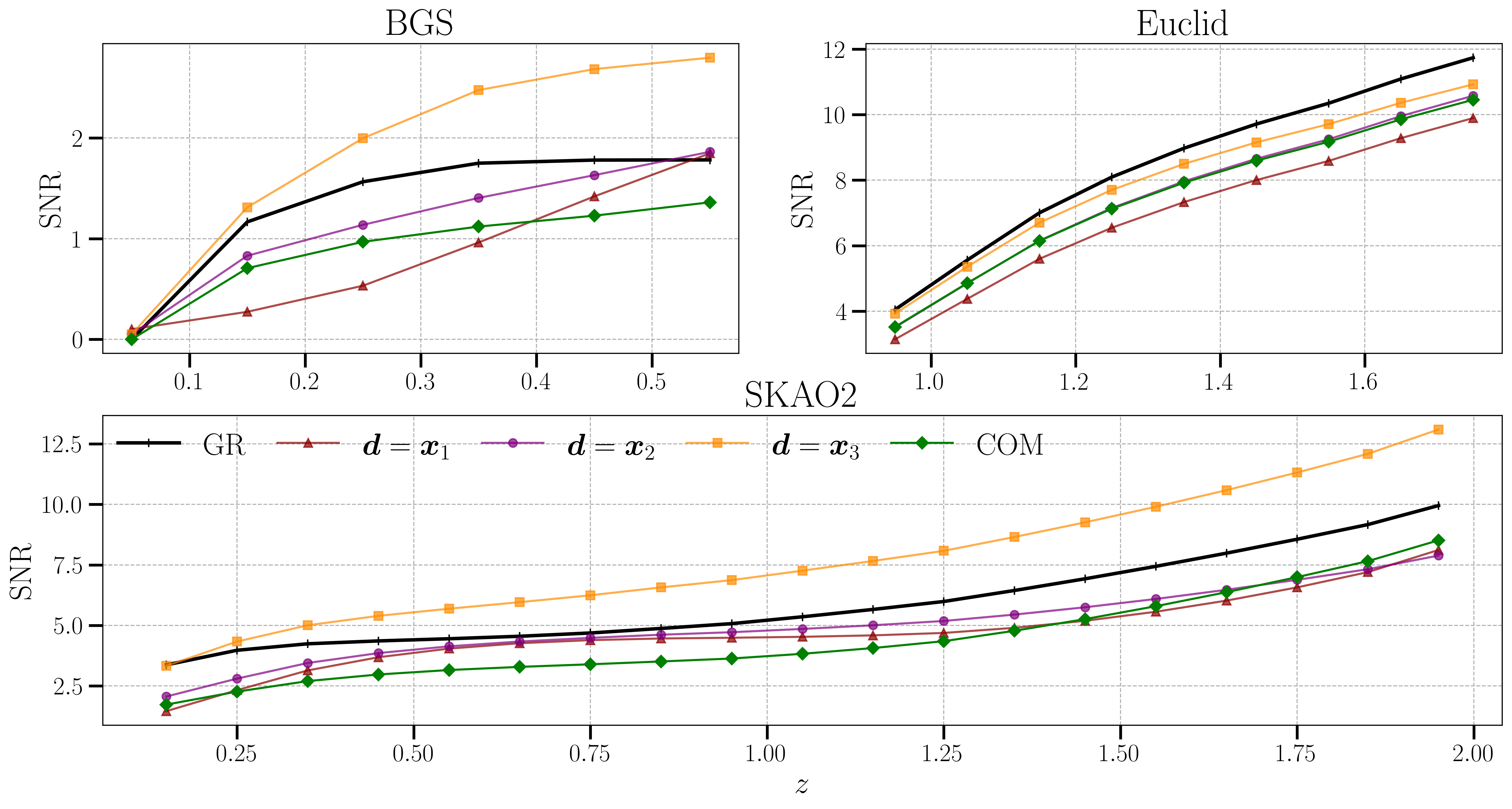}
    \caption{\label{fig:SNR} Cumulative SNR of the dipole, $(\ell=1,m=0)$ plotted over redshift bins, for the three different surveys we consider, without wide-separation effects (\textit{green}) -- that is, purely relativistic contribution -- and with wide-separation for different LOS choices.}
    \end{figure}

    We compute the SNR in redshift bins of width $\Delta z =0.1$ and then the cumulative SNR for a particular survey is given by summing the SNR from each bin in quadrature ${\rm SNR}(<z) = \sum_z \sqrt{{\rm SNR}(z)^2}$. In $k$-space, we impose bin widths of $\Delta k = 2 k_f$ where $k_f$ is the fundamental frequency of the survey ($k_f \approx (2 \pi)/V^{1/3}$) and a cut-off scale at $k_{max} =0.1\: [h/ {\rm Mpc}]$. We also impose an effective $k_{\rm min}$ by excluding Fourier modes where $k > 2 \pi/x(z_{\rm min})$ for the minimum redshift included in the bin, as the wide-angle expansion breaks down (see discussion at the end of Section \ref{Sec:geom_config}). Results are plotted in Figure~\ref{fig:SNR} and SNR values are given in Table~\ref{tab:SNR}.

    \begin{table}[tbp]
    \centering
    \begin{tabular}{|c|c|c|c|c|}
    \hline
    &DESI BGS & Euclid H$\alpha$ & SKAO1 & SKAO2\\
    \hline
    ${\rm SNR}_{\rm WA}$& 1.9 & 1.9 & 1.2 & 5.5 \\
    ${\rm SNR}_{\rm RR}$& 0.3 & 2.4 & 0.1 & 2.5 \\
    ${\rm SNR}_{\rm WS}$& 1.6 & 1.1 & 1.2 & 5.9 \\
    ${\rm SNR}_{\rm GR}$& 1.8 & 11.7 & 0.4 & 8.8 \\
    ${\rm SNR}_{\rm All}$& 2.8 & 10.9 & 1.3 & 13.1 \\
    \hline
    \end{tabular}
    \caption{\label{tab:SNR} SNR for the dipole, $\ell=1,m=0$, of the galaxy bispectrum for different contributions for each survey we consider. Wide-separation terms (WS) includes both the wide-angle and the radial evolution contributions.}
    \end{table}

    We note that this analysis ignores several key factors and just constitutes a crude calculation. Firstly, these results will be heavily impacted by the convolution with the survey window function which will dampen the signal on the large scales as well as mixing the odd and even parity signals. On smaller, mildly nonlinear scales, the tree-level theory assumed here will be inadequate both in the signal and the covariance. In general, much more attention needs to paid to the covariance; the Gaussian covariance limit has been shown to be a poor approximation to the true covariance, particularly in the squeezed limit \cite{Barreira_2019,Biagetti_2022,Salvalaggio_2024}, and also we have neglected the relativistic and wide-separation contributions to the covariance. Lastly, proper consideration of small scales should allow for higher $k_{\rm max}$ allowing a greater range of modes to be accessed, rather than the harsh cut-off of $k_{\rm max}=0.1\: [h/ {\rm Mpc}]$ considered here. Our results do however appear to be broadly consistent with those of previous analyses \cite{Maartens_2020,Noorikuhani_2022,Rossiter_2024}.

\begin{figure}[tbp]
        \centering 
        \includegraphics[width=0.75\textwidth]{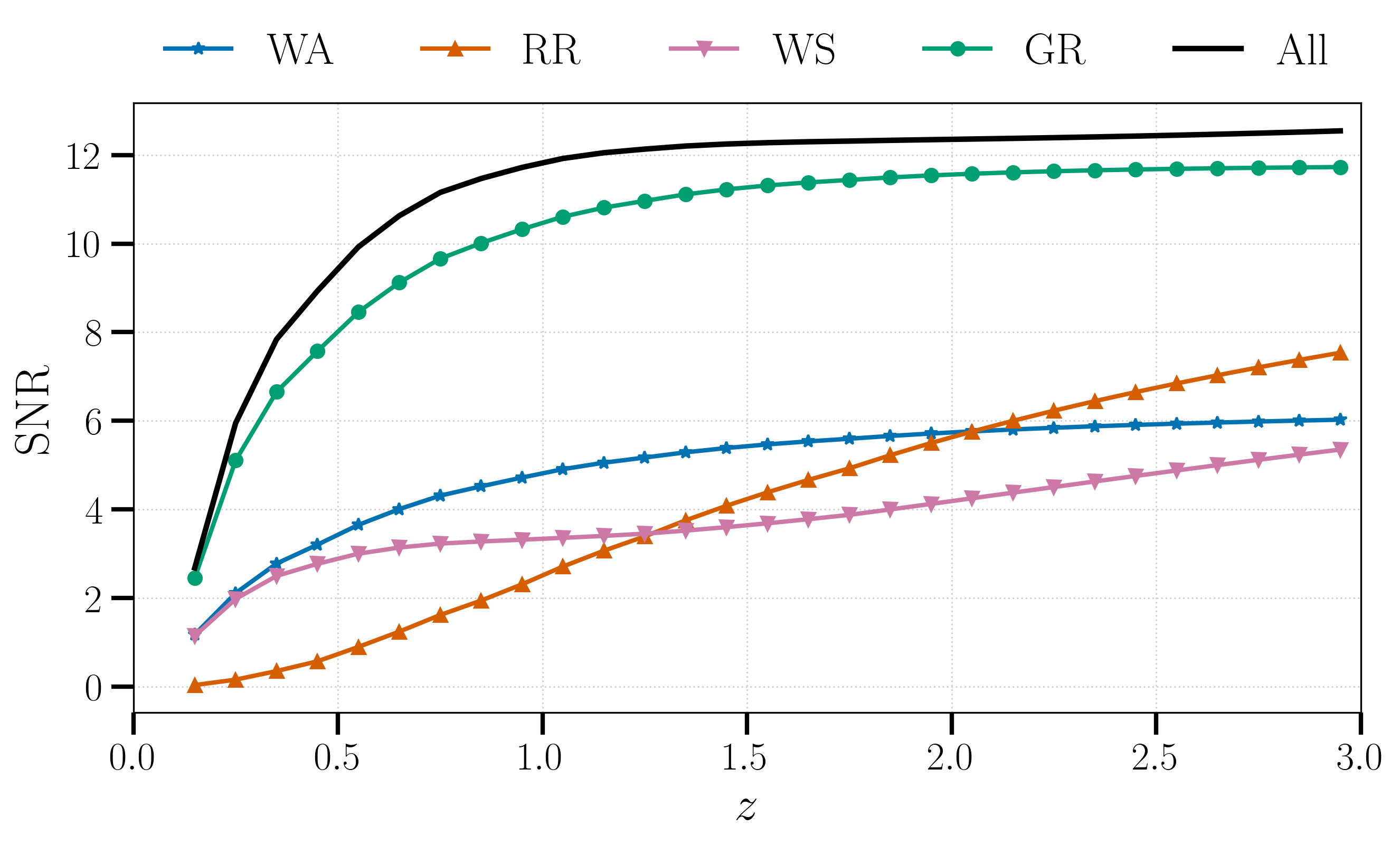}\caption{\label{fig:SNR_CVlim} Cumulative SNR for different contributions to the dipole of the bispectrum for a cosmic variance-limited survey with $b_e=\mathcal{Q}=0$. Also $b_1 = \sqrt{1+z}$ and $b_2$ and $b_{\Gamma_2}$ are set by Equation~\eqref{eq:secondorderbias}.}
    \end{figure}

    The SNR for this type of forecast is strongly dependent on survey volume, as larger volumes lead to a smaller $k_f$ and, therefore, accessing a greater number of triangles. The predominant part of the signal arises from squeezed, or moderately squeezed triangles, where $k_3$ is small and $k_1$ and $k_2$ are comparatively large. The low redshift samples of DESI BGS and SKAO1-like samples do not cover a large enough volume for a strong detection of the relativistic signal while for the larger redshift coverage of SKAO2 and Euclid, we obtain ${\rm SNR} \sim \mathcal{O}(10)$. Therefore, one would expect the relativistic terms to be detectable in other similar spectroscopic surveys, like the higher redshift ELG and LRG samples of DESI. For a cosmic variance-limited survey, with $f_{\rm sky}=1$ and assuming $b_e=\mathcal{Q}=0$, as shown in Figure~\ref{fig:SNR_CVlim}, also has a  relativistic signal of ${\rm SNR} \sim \mathcal{O}(10)$. The dominant non-evolution and magnification bias terms generally are inversely proportional to redshift, however for a realistic survey, non-zero evolution and magnification biases will drive the relativistic signal at higher redshifts.  

    Therefore, for a realistic analysis, as well as accessing more modes by pushing to higher redshifts, greater constraining power on relativistic corrections should be achievable with multi-tracer approaches (e.g. see \cite{Karagiannis_2024} for an example of constraints from a multi-tracer analysis); indeed, the leading $(\mathcal{H}/k)$ relativistic corrections enter the odd multipoles of the power spectrum in the multi-tracer case. 
    
    Further, while we just considered the dipole, $\ell=1,m=0$, for the discussion here, additional information from the imaginary bispectrum is contained in the other odd multipoles. 

    Confident detection of the wide-separation corrections requires large volumes, but as expected the wide angle signal peaks at lower redshifts while the radial redshift signal is fairly redshift independent. Their effect on the overall SNR of the dipole, as shown in figure \ref{fig:SNR}, is survey dependent and is also dependent on the LOS choice.
    
    If we consider the linear-order expansions in Equations \eqref{eq:WA_expansion} and \eqref{eq:RR_expansion}, then the wide-angle terms contain either the dot products $\bs{d}\cdot\hat{\bs{x}}_{1i}$, or $\bs{k}\cdot\hat{\bs{x}}_{1i}$, and the radial-redshift terms have a factor of $\bs{d}\cdot\hat{\bs{x}}_{1i}$. In general, these $\bs{d}\cdot\hat{\bs{x}}_{1i}$ terms appear more suppressed for a less extreme LOS choice; often, as in the Euclid-like case, the $\bs{d}\cdot\hat{\bs{x}}_{1i}$ and $\bs{k}\cdot\hat{\bs{x}}_{1i}$ contributions are of opposite sign and, therefore, the wide-angle contributions are larger for a COM LOS as there is less cancellation between the two contributions.

\begin{figure}[tbp]
    \centering 
    \includegraphics[width=0.75\textwidth]{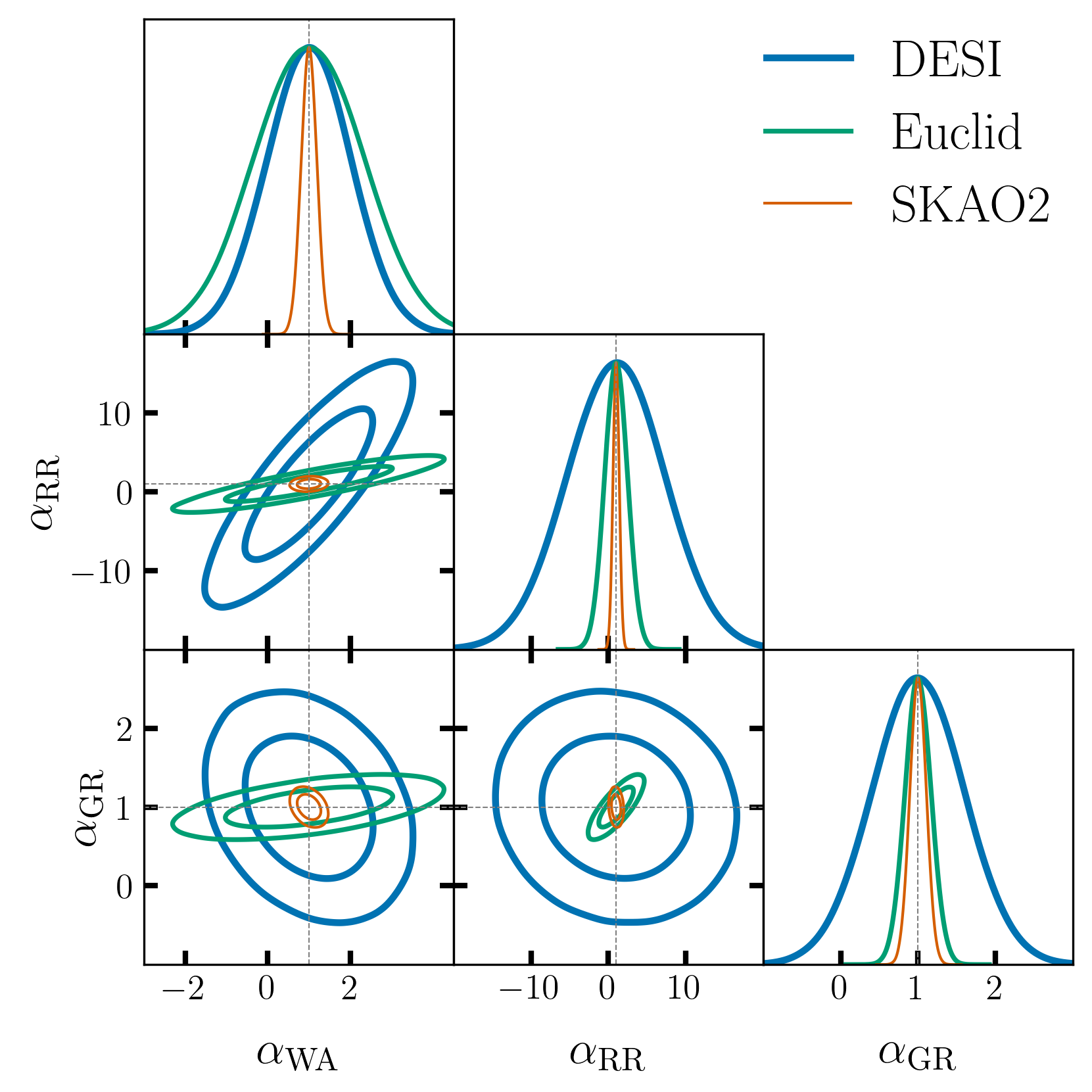}
    \caption{\label{fig:fisher} 1- and 2-$\sigma$ forecast for the amplitude parameters measured with the dipole of the bispectrum for a DESI BGS (\textit{thick blue}), Euclid H$\alpha$ (\textit{green}) and a SKAO2 HI (\textit{thin orange}) galaxy surveys.}
    \end{figure}

    \paragraph{Fisher forecast}
    If we introduce amplitude parameters for each effect ($\alpha_{\rm GR},\alpha_{\rm WA},\alpha_{\rm RR}$), such that the odd part of the bispectrum is given by
    \begin{equation}
        B^{(1)}_{\rm loc}= \alpha_{\rm WA}B_{\rm WA_1}+\alpha_{\rm RR}B_{\rm RR_1}+\alpha_{\rm GR}B_{\rm GR_1},
    \end{equation}
    we can examine the degeneracy between the contributions by performing a Fisher matrix analysis on these parameters. Assuming a Gaussian likelihood, the Fisher matrix depends on the derivative of the multipoles with respect to the parameters:
    \begin{equation}
        F_{ij,\ell m} = \sum_{\triangle} \frac{ \partial B_{\ell m}}{\partial \theta_i}{\sf C}^{-1}_{\ell m}\frac{ \partial B_{\ell m}^*}{\partial \theta_j}.
    \end{equation}
    Figure~\ref{fig:fisher} shows the 1- and 2-$\sigma$ forecasted constraints on the amplitude parameters for the DESI-, Euclid-, and SKAO2-like surveys. For the Euclid case, the wide-separation and relativistic contributions are positively correlated, unlike in the SKAO2-like galaxy survey like case. We can see the shift in the correlations due to different ranges of redshift for each survey. 
    \subsection{Parity even}

    The parity even part of the wide-separation and relativistic corrections, which comes in at second order, includes a leading relativistic and first-order wide-separation mixing contribution, such that we can write
    \begin{equation}
        B^{(2),{\rm loc}} = \left(\frac{i}{k_i \, d}\right)^2 (B_{{\rm WS}_2}) + \frac{i\,{\cal H}}{k_i}\left(\frac{i}{k_i \, d}\right)(B_{{\rm WS}_1 {\rm GR}_1}) + \left(\frac{i\,{\cal H}}{k_i}\right)^2 B_{{\rm GR}_2},
    \end{equation}
    where WS includes both wide-angle and radial-redshift contributions, as well as the mixing between the two leading order terms.
    
\begin{figure}[tbp]
    \centering 
    \includegraphics[width=\textwidth]{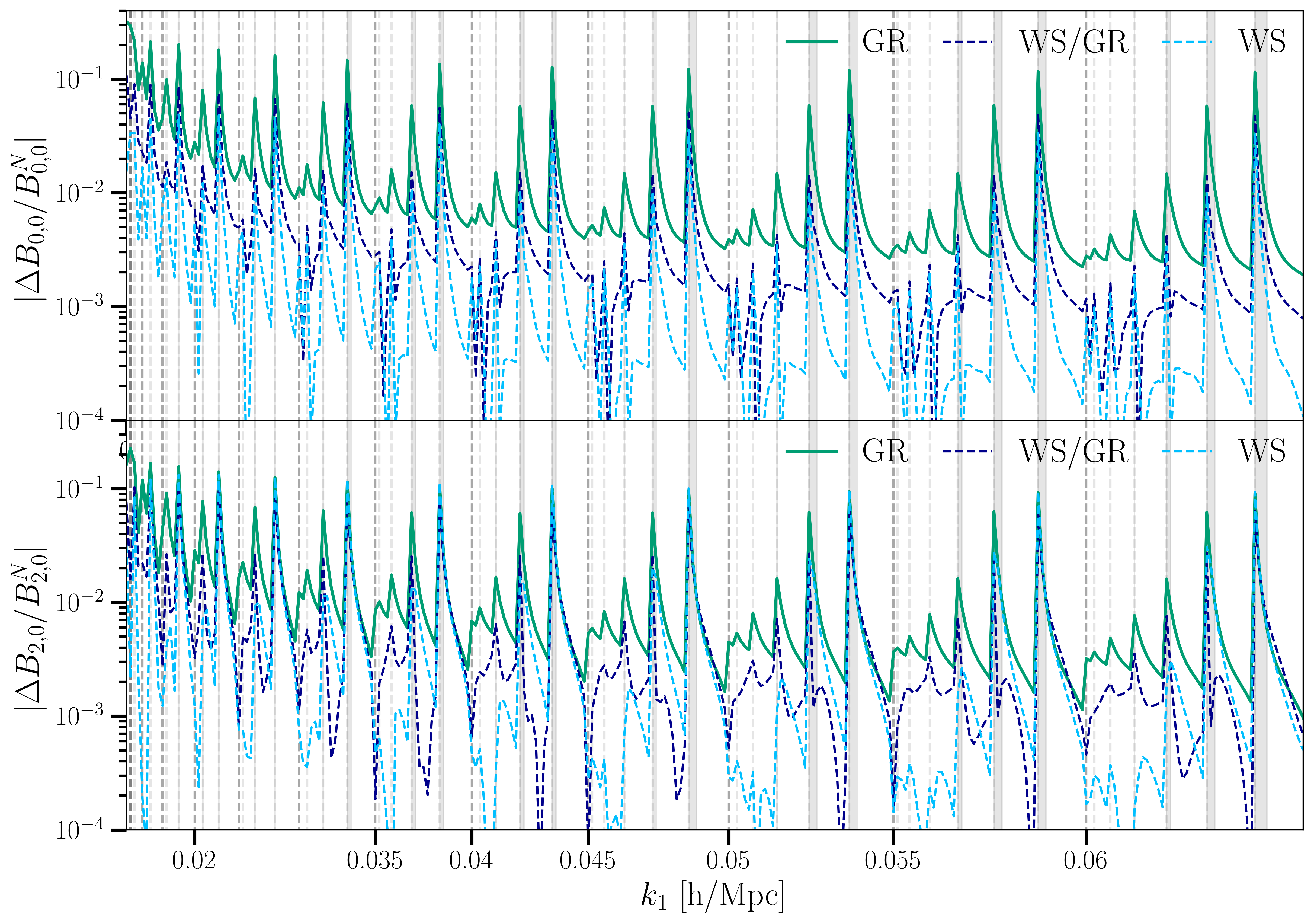}
    \caption{\label{fig:secondorder_alltri} Fractional contribution to the monopole (\textit{top}) and quadrupole (\textit{bottom}) from second order wide-separation (WS), relativistic effects (GR) and the mixing contribution (WS/GR) for a H$\alpha$ Euclid-like survey at $z=1$, plotted over all triangles. See Figure~\ref{fig:dipole_all} for a full description of the vertical lines.}
    \end{figure}

\begin{figure}[tbp]
    \centering 
    \includegraphics[width=\textwidth]{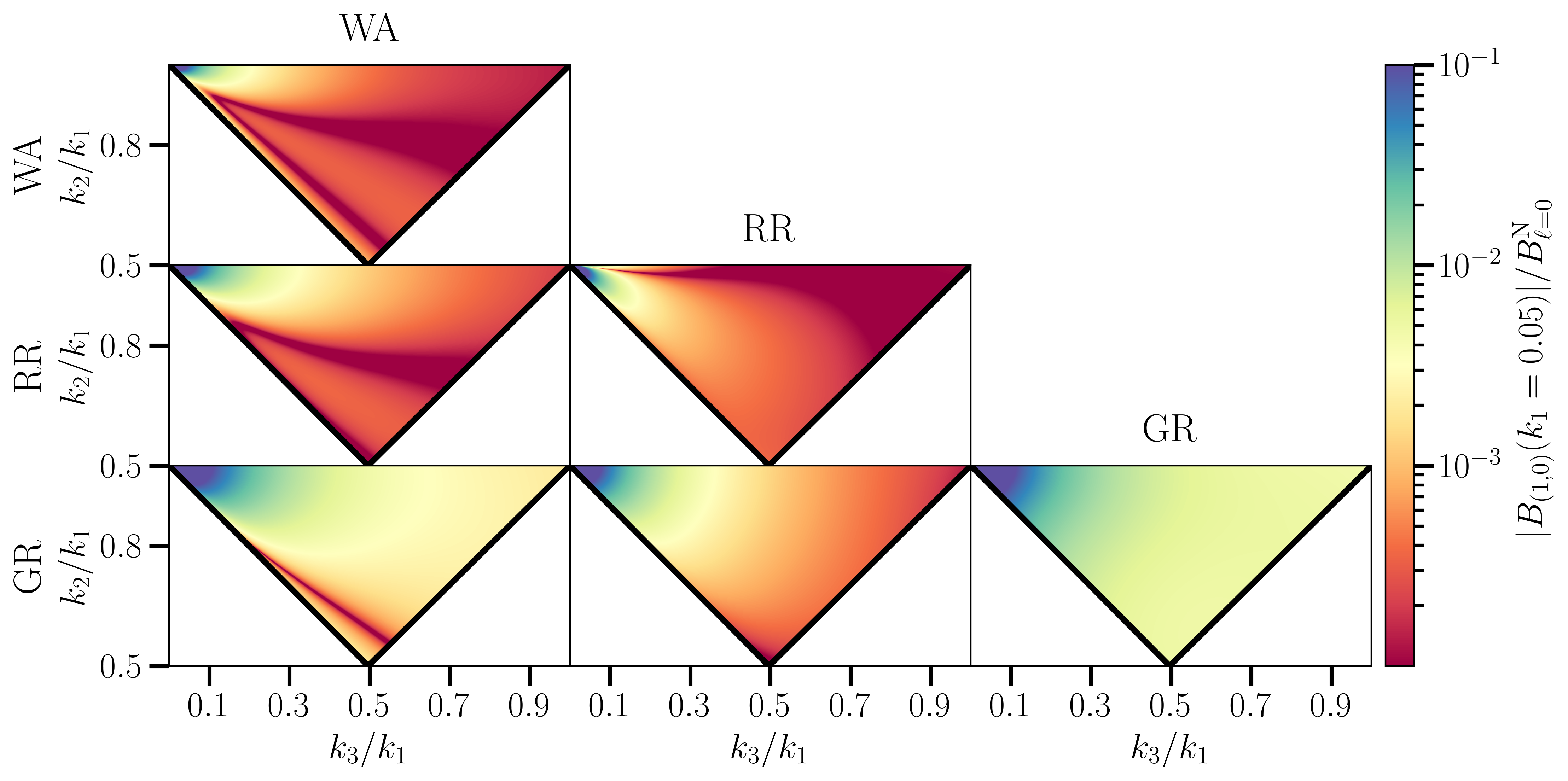}
    \caption{\label{fig:even_terms} Fractional contribution to the monopole from second order wide-separation and relativistic effects for a H$\alpha$ Euclid-like survey at $z=1$. This is plotted over bispectrum shapes, for a fixed scale with $k_1=0.05 \, \, [h/Mpc]$. For each triangle: Top left corner: Squeezed, Top right corner: Equilateral, Bottom: Folded. }
    \end{figure}

\begin{figure}[tbp]
    \centering 
    \includegraphics[width=\textwidth]{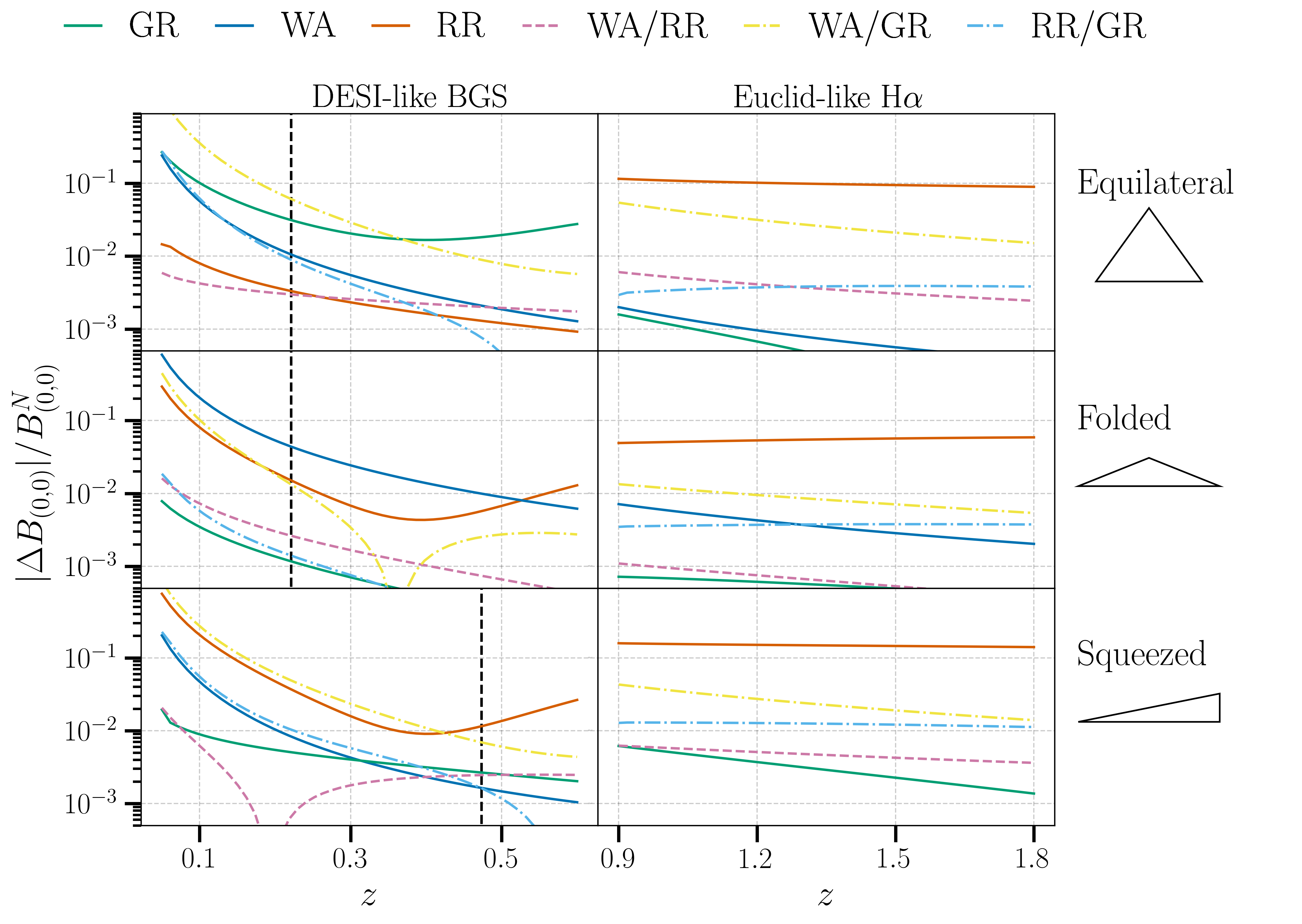}
    \caption{\label{fig:redshift_second} Even parity second-order contributions to the monopole as a fraction of the Newtonian part for three different triangle configurations (equilateral, squeezed and folded), plotted as a function of redshift, for both a DESI BGS and an H$\alpha$ Euclid-like survey.}
    \end{figure}

    As in the case of the odd-parity terms, the even-parity relativistic contribution, $({\rm GR}_2)$, is generally larger than the wide-separation contributions (as in Figures~\ref{fig:secondorder_alltri},~\ref{fig:even_terms} and~\ref{fig:redshift_second}), $({\rm WA}_2 + {\rm RR}_2 +{\rm WA /RR})$ over all shapes and scales; though at second order we also have the mixing contribution arising from the wide-separation corrections to the leading-order relativistic terms, and this contribution is in general more important than the pure wide-separation contribution.

    These real, even, second-order contributions are smaller than their imaginary, odd, first-order counterparts. The percentage correction of wide-separation effects (including the mixed terms) to the standard Newtonian term is $<1 \%$ for most triangles (e.g. see Figure~\ref{fig:secondorder_alltri}), but it peaks on small scales and in the squeezed limit, where it can be of order $10 \%$.

    The general shape dependence (e.g. Figure~\ref{fig:even_terms}) and redshift dependence (see Figure~\ref{fig:redshift_second}) is similar to the first order contributions, with the most notable features being the $1/k^2$ dependence for all contributions and the $1/\chi(z)^2$ dependence for the wide-angle terms.
    
    \subsubsection{Effective PNG}
    While the percentage contributions of wide-separation corrections to the monopole are of order $<1 \%$ for scales $k\approx 0.01 \, [h/ {\rm Mpc}]$, with a larger contribution coming from the mixing with the relativistic corrections, it is important to consider for an accurate analysis of PNG; if wide-separation and relativistic corrections are ignored in the theory modelling then this unaccounted for signal can mimic a PNG signal. Previous studies have shown that relativistic corrections can significantly bias measurements for PNG of the local type in the monopole of the galaxy bispectrum \cite{Umeh_2017,Dio_2017}, and here we consider a similar analysis with the inclusion of wide-separation.
    
    \begin{figure}[tbp]
    \centering 
    \includegraphics[width=\textwidth]{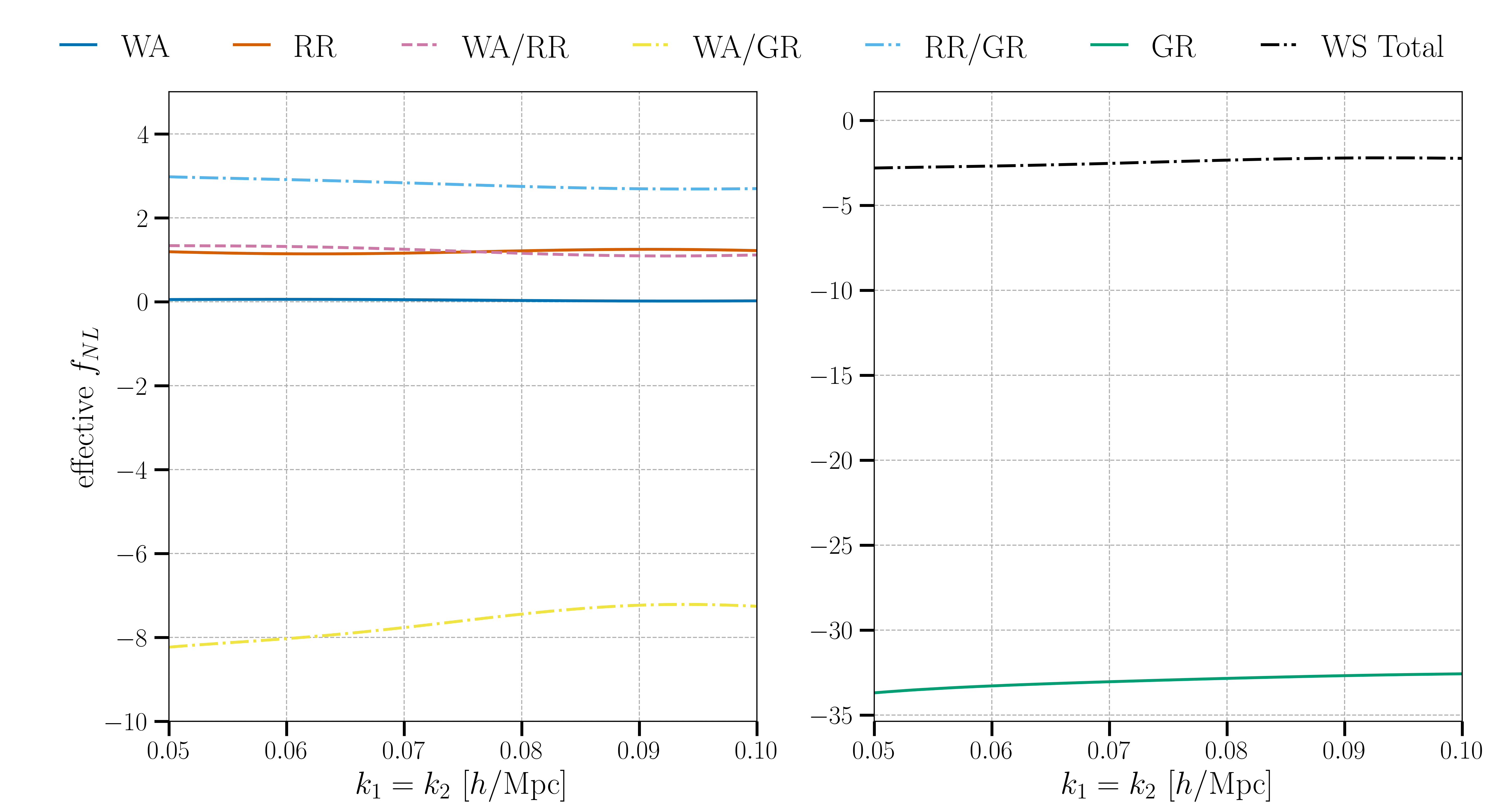}
    \caption{\label{fig:effective_fnl} Left Panel: Effective local $\fNL$ induced in the monopole by each wide-separation contribution (including mixing with relativistic terms) for a squeezed limit configuration with $k_3=0.01$, for a Euclid-like survey at $z=1$. A COM LOS is used for the monopole. Right Panel: Combined effective $\fNL$ for all wide-separation effects alongside the pure relativistic contribution.}
    \end{figure}
    
    Focusing on local type PNG and as such we consider the squeezed limit of the bispectrum monopole at large scales. We consider a local PNG contribution including scale dependent bias following \cite{tellarini2016}, with full details of our modelling given in Appendix \ref{ap:local_PNG}. The left panel of Figure~\ref{fig:effective_fnl} shows the effective local $f_{\rm NL}$ induced by each type of wide-separation correction, for the Euclid-like bias parameters at $z=1$. We can see that while the pure wide angle contribution mimics $f_{\rm NL}$ of $\mathcal{O}(0.1)$, which is consistent with the findings in \cite{Pardede_2023}, the bias from the other effects is more significant. The relativistic/wide-separation mixing in particular can lead to the individual contributions with $f_{\rm NL} \approx 10$. The right panel 
    shows the total wide-separation corrections, including all mixing terms, as well as the dominant pure relativistic term.

    In comparison to previous results, the effective local non-Gaussianity induced by GR effects, $f^{\rm eff \, GR}_{\rm NL}$, is larger than that in \cite{Umeh_2017,Dio_2017}. This can attribute this largely to the signal from the evolution and magnification bias terms. For $b_{\rm e}=\mathcal{Q}=0$, we find $f^{\rm eff \, GR}_{\rm NL}\approx 5$ roughly in line with the results of \cite{Umeh_2017}, who considered the impact of local relativistic projection effects on the full bispectrum. Detailed comparison with the results of \cite{Dio_2017} is not straightforward, as they consider the bispectrum in a spherical basis. For a higher redshift sample, the wide-angle contribution will decrease (see Figure~\ref{fig:redshift_second}), but the radial-redshift contribution, in particular, will still be relevant for precision constraints. 

    We note that these results are for this specific redshift, triangle configuration, and scale as well as for our chosen bias models and we stress that while it can mimic a sizeable $\fNL$ for this configuration, we would not necessarily expect this if one were to sum over all triangles. We therefore plan to return to study the impact of these effects of the measurements of PNG in a future work.
    
    Wide-separation corrections directly to the PNG contribution (wide-separation/PNG mixed term) are negligible for most analyses; full relativistic treatment of PNG is more subtle and has been studied in \cite{Maartens_2021}. 
    
    The effective $f_{\rm NL}$ induced by the wide-separation corrections in the galaxy bispectrum falls within current constraints for local-type $\fNL$ \cite{Cabass_2022,Damico_2022} \footnote{These constraints come from joint power spectrum and bispectrum, with constraints tightening by $20\%$ with the inclusion of the bispectrum. For constraints on non-local shapes of $\fNL$ however, the bispectrum is necessary \cite{Cabass_2022_single}.}, however to reach the goal of $\Delta f_{\rm NL}\approx 1$, both wide-separation and relativistic corrections cannot be ignored.
    
\section{Conclusions}\label{sec:conclusions}

In this work, we computed wide-separation effects to the three-dimensional galaxy bispectrum with a generalised line-of-sight orientation in the triplet of galaxies, including, for the first time, the contribution from local redshift evolution, which becomes relevant when galaxies are widely separated in terms of radial distance. The expressions and routines used to compute these effects for given kernels of the galaxy bispectrum are publicly available \href{https://github.com/craddis1/ws_bk_theory}{\faGithub}. For low redshifts and large angular footprints, the perturbative approach to wide-angle effects breaks down at the largest scales. However, this perturbative approach should be robust enough to make precision constraints for the most relevant scales in ongoing (DESI and Euclid) and forthcoming (SKAO2) surveys that map large redshifts.

We show that the imaginary part of these effects, which enter the odd multipole moments of the bispectrum, can be up to 10\% of the Newtonian monopole for ongoing Stage-IV galaxy surveys, like Euclid (see Figure \ref{fig:dipole_all}). At lower redshifts, as in a DESI-like Bright Galaxy Survey (BGS), wide-angle effects (WA) are the dominant part of the wide-separation correction, but for $z>1$ both the WA and RR effects are of a similar magnitude as in the Euclid-like case considered here (Figure \ref{fig:redshift_first}). We stress, nonetheless, the dependence of our results on the range of scales, shapes, redshifts and bias parameters considered.

We have compared these contributions to the signal from general relativistic corrections, including both dynamical and projection effects. These contributions are, in general, larger than the wide-separation terms for the cases considered here, and the leading-order imaginary part that enters the odd multipoles should be detectable with a single tracer in surveys with large enough volume: for a Euclid-like H$\alpha$ galaxy survey, over $0.9<z<1.8$, our forecast showed a signal-to-noise ratio of order 10. Since the odd multipoles of the galaxy bispectrum are a possible pathway to constrain gravity on cosmological scales, this work has highlighted that it is important to accurately model and account for the wide-separation effects -- including the radial evolution contribution -- due to their degeneracies with the relativistic terms. 

The second-order wide-separation effects are real and affect even-parity multipoles. Here, we showed that these can mix with the relativistic signal, and that the mixing terms between wide-separation and relativistic effects can be of similar order to the pure $(\mathcal{H}/{k})^2$ relativistic signal. These effects, while constituting percentage or sub-percentage level corrections to the Newtonian contribution, will still need to be considered for precision analysis on large scales ($k\lesssim 10^{-2}\,\,h/{\rm Mpc}$), which are particularly relevant to constrain PNG. For PNG of the local-type, we have showed that wide-separation effects in the squeezed limit, if unaccounted for, can `mimic' $\fNL$ up to $\mathcal{O}(10)$ (Figure \ref{fig:effective_fnl}) mainly through mixing with the relativistic terms. The effective $\fNL$ we find generated by the pure relativistic term alone, $\fNL \approx - 30$, is larger than in previous analysis, but this result is not only specific to this particular configuration, scale and redshift etc, but also can be attributed to the models of evolution and magnification bias we use. In future work we plan to consider more detailed forecasts of the bias induced on the best fit measurement of PNG if these systematics are ignored for the most common PNG shapes.

In this work, we omitted the effect of the convolution of the survey window function which significantly impact the large scales analysed, and therefore it is important to model how these contributions will be affected. The modelling of this is non-trivial, and we leave this aspect as an avenue of future work. Further, the impact of nonlinear linear effects should be considered, and the inclusion of mode-coupling, off-diagonal and beyond-Gaussian terms should be included in a more realistic modelling of the covariance matrix.

\acknowledgments
CA thanks Stefano Zazzera for useful discussions. CA is supported by a studentship from the UK Science and Technology Facilities Council (STFC). CG and CC acknowledge financial support from the UK STFC consolidated grant ST/T000341/1. This work made extensive use of the public code \href{https://github.com/lesgourg/class_public}{\sc{class}} \cite{Lesgourgues2011,blas2011}, and the following \texttt{python} packages and libraries: \href{https://numpy.org/}{\sc{numpy}} \cite{harris2020}, \href{https://scipy.org/}{\sc{scipy}} \cite{2020SciPy} and \href{https://matplotlib.org/}{\sc{matplotlib}} \cite{hunter2007}.

\section*{Code Availability Statement}
\textsc{CosmoWAP} is publicly available at \url{https://github.com/craddis1/CosmoWAP}.

\appendix

\section{Impact of the survey window convolution}\label{ap:window_convolution}

The window function convolution will dampen the signal for $k$-modes that approach the size of the survey.  Additionally, from the convolution, for a given term with $\mu^m$ dependence it introduces additional terms dependent on $\mu^{m-n}/k^n$. The effect of this is to mix the parity odd and even terms such that different signals enter each multipole, though this is suppressed on scales much smaller than the window.

For completeness, we briefly examine the effect of the survey window on the bispectrum, but implementation is beyond the scope of this work (see \cite{Sugiyama_2018,Pardede_2022,Wang_2024} for more detailed discussions).

We can define our local windowed bispectrum 
\begin{equation}
    B^{\rm W}_{\rm loc}(\bs{k}_{1},\bs{k}_{2},\bs{d}) = \int_{\bs{x}_{13},\bs{x}_{23}}\,{\rm e}^{-i(\bs{k}_1 \cdot \bs{x}_{13} + \bs{k}_2 \cdot \bs{x}_{23})}\,W(\bs{x}_1)W(\bs{x}_2)W(\bs{x}_3)\, \zeta_{\rm loc}(\bs{x}_{13},\bs{x}_{23},\bs{d})
\end{equation}
from our theoretical unwindowed expression with wide-separation corrections already included in $\zeta_{\rm loc}(\bs{x}_{13},\bs{x}_{23},\bs{d})$. Note that here wide-separation corrections are computed directly on the unwindowed theory and not on the full windowed expression.

If we assume $\bs{d}=\bs{x}_3$ ($\zeta_{\rm loc}(\bs{x}_{13},\bs{x}_{23},\bs{d})$ is computed for the same LOS) then one can write:

\begin{equation}
    \begin{aligned}
        B^{W}_{\rm loc}(\bs{k}_{1},\bs{k}_{2},\bs{d}) \equiv W(\bs{d}) \int_{\bs{x}_{13},\bs{x}_{23}} \,{\rm e}^{-i(\bs{k}_1 \cdot \bs{x}_{13} + \bs{k}_2 \cdot \bs{x}_{23})}\,W(\bs{d}+\bs{x}_{13})W(\bs{d}+\bs{x}_{23})\zeta_{\rm loc}(\bs{x}_{13},\bs{x}_{23},\bs{d}).
    \end{aligned}
\end{equation}
By defining the Fourier transform of the window
\begin{equation}
    W(\bs{x}) =\int \dd^3 \bs{q} \, W(\bs{q}) e^{i \bs{q}\cdot {x}},
\end{equation}
and writing the local correlation as the inverse Fourier transform of the local Fourier bispectrum, then the windowed bispectrum is given by
\begin{equation}
    \begin{aligned}
        B^{W}_{\rm loc}(\bs{k}_{1},\bs{k}_{2},\bs{d}) \equiv W(\bs{d})\int_{\bs{x}_{13},\bs{x}_{23},\bs{k}_{1}',\bs{k}_{2}',\bs{q}_1,\bs{q}_2} \,{\rm e}^{-i(\bs{k}_1 - \bs{k}_1' +\bs{q}_1)\cdot \bs{x}_{13}}{\rm e}^{-i(\bs{k}_2 - \bs{k}_2' +\bs{q}_2)\cdot \bs{x}_{23}}\,\\ \times W(\bs{q}_1)W(\bs{q}_2)B_{\rm loc}(\bs{k}_{1},\bs{k}_{2},\bs{d}),
    \end{aligned}
\end{equation}
The configuration space integrals then become Dirac deltas, which contract the integrals over $\bs{k}_{1}',\bs{k}_{2}'$ such that the convolution of the local bispectrum with window becomes
\begin{equation}
        B^{W}_{\rm loc}(\bs{k}_{1},\bs{k}_{2},\bs{d}) \equiv W(\bs{d})\int_{\bs{q}_1,\bs{q}_2} \,W(\bs{q}_1)W(\bs{q}_2)B_{\rm loc}(\bs{k}_{1}+\bs{q}_1,\bs{k}_{2}+\bs{q}_2,\bs{d}).
\end{equation}

\section{Gaussian covariance for the bispectrum multipoles}\label{ap:covariance}

The estimator defined in Equation \eqref{eq:estimator}, assuming the plane-parallel limit and ignoring the convolution of the survey window on the multipoles, simplifies to
\begin{equation}
     \hat{B}_{\ell,m}(k_1,k_2,k_3) = \frac{1}{V_{123}}\int_{\mathcal{S}_1} \dd^3 \bs{q}_1 \int_{\mathcal{S}_2} \dd^3 \bs{q}_2 \int_{\mathcal{S}_3} \dd^3 \bs{q}_3 \delta^D(\bs{q}_{123}) \delta(\bs{q}_1)\delta(\bs{q}_2)\delta(\bs{q}_3) Y^*_{\ell,m}(\hat{\bs{q}}_1 \cdot \hat{\bs{d}},\phi).
\end{equation}

The covariance of each multipole is given by
\begin{equation}
    {\sf C}^B_{\ell m}(k_1,k_2,k_3,k'_1,k'_2,k'_3) \equiv \langle \hat{B}_{\ell m}(k_1,k_2,k_3) \hat{B}^*_{\ell m}(k'_1,k'_2,k'_3)\rangle-\langle \hat{B}_{\ell m}(k_1,k_2,k_3)\rangle \langle \hat{B}^*_{\ell m}(k'_1,k'_2,k'_3)\rangle.
\end{equation}

To model the covariance for the bispectrum multipoles we make the simplifying assumptions of assuming Gaussian cosmic variance and only considering the leading order Newtonian terms. For Gaussianity $\langle \hat{B}_{\ell m}(k_1,k_2,k_3)\rangle$ is zero and as such following \cite{Chan_2017} we can write:
\begin{equation}
    {\sf C}^B_{\ell m} = \frac{s_B}{4 \pi \, V_{123}}  \int \dd \mu_1 \int \dd \phi \, |Y_{\ell}^m(\mu_1,\phi)|^2 P_{\rm PP}(k_1,\mu_1)P_{\rm PP}(k_2,\mu_2)P_{\rm PP}(k_3,\mu_3)
\end{equation}
where we assume the thin bin limit $\Delta k << k$ such that $V_{123}= 8 \pi^2 k_1 k_2 k_3 (\Delta k /k_f)^{3}\beta,$ and the fundamental frequency of the survey is given by $k_f \approx 2 \pi/(V)^{(1/3)}$. Additionally, $\beta=1$ except for the case where $k_1 = k_2 + k_3$ where $\beta= 1/2$, $s_{123}=1$ for scalene, $s_{123}=2$ for isosceles and $s_{123}=6$ for equilateral triangles. Finally, $P_{\rm N}(k_i,\mu_i)$ is the standard kaiser plane-parallel redshift space power spectrum with a shot noise contribution to the noise given by 
\begin{equation}
    P_{\rm N,PP+N}(k,\mu)= Z_{\rm N}^{(1)}(k,\mu)^2 P_{\rm lin}(k) + \frac{1}{ \bar{n}_g}.
\end{equation}

\section{Other multipoles}\label{ap:other_multipoles}

\begin{figure}[tbp]
    \centering 
    \includegraphics[width=\textwidth]{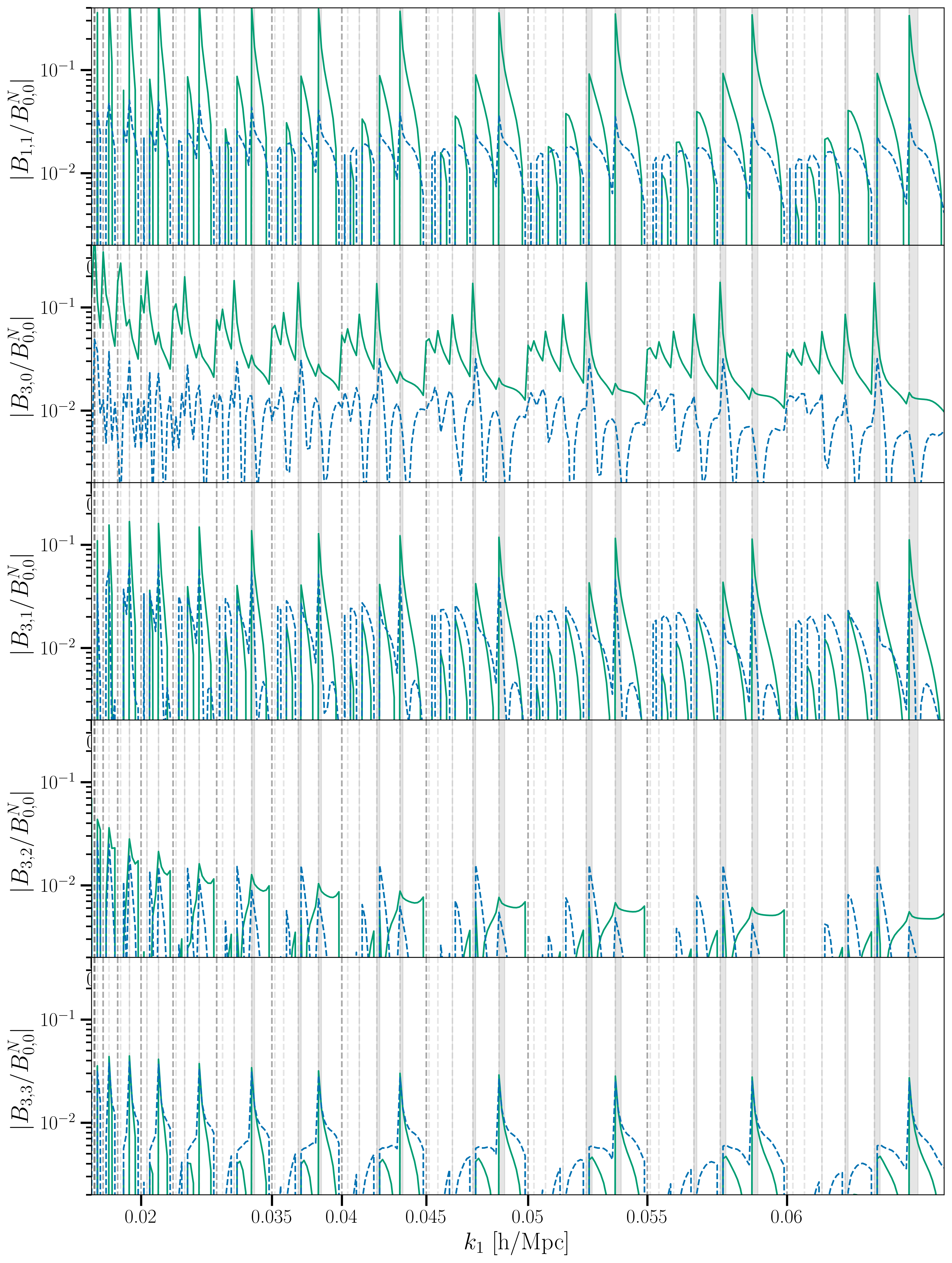}
    \caption{\label{fig:other_odd} Relativistic (\textit{green}) and wide-separation (\textit{blue}) corrections to a selection of odd multipoles. These contain additional information to that included in the dipole alone.}
\end{figure}

Figure~\ref{fig:other_odd} shows the contribution from the imaginary bispectrum, generated by wide-separation and relativistic corrections, for a selection of odd multipoles. Without wide-separation corrections, multipoles up to $\ell=6$ are induced. However, for the $n^{\rm th}$ order in the wide-separation expansion, non-zero multipoles are generated up to $\ell = 6 + n$.

In comparison to Figure~\ref{fig:dipole_all} we can see different shape dependence for the $m\neq0$ multipoles; in particular the relativistic contributions appears to have a distinct peak in the squeezed isosceles triangles. Note, though the amplitude of the multipoles is smaller for higher $m$, evaluation of the additional information from each multipole requires further analysis, in particular the cross-multipole covariance and is something we may return to in the future.
    
\section{Newtonian Local type Non-Gaussianity}\label{ap:local_PNG}

We consider the local non-Gaussianity contribution, including scale-dependent biases, following \cite{tellarini2016}, the non-Gaussian contributions to the first-order and second-order perturbations theory kernels are given by
\begin{equation}
    Z^{(1)}_{\rm N,NG} = D(x_1)  b_{01}(x_1)
\end{equation}
and
\begin{equation}
\begin{aligned}
    Z^{(2)}_{\rm N,NG} = \, D^2(x_1) & \left[b_1(x_1) \left(\fNL \frac{\alpha(q_3)}{\alpha(q_1)\alpha(q_2)}\right) + f(x_1)\frac{(\bs{q}_{12}\cdot \hat{\bs{x}}_1)^2}{q^2_{12}}\left(\fNL \frac{\alpha(q_3)}{\alpha(q_1)\alpha(q_2)}\right)\right.  \\ & \, \left. \frac{b_{11}(x_1)}{2}\left(\frac{1}{\alpha(q_1)}+\frac{1}{\alpha(q_2)}\right) -  b_{01}(x_1)\left(\frac{(\bs{q}_1\cdot\bs{q}_2)}{q_1^2\alpha(q_2)}+\frac{(\bs{q}_1\cdot\bs{q}_2)}{q_2^2\alpha(q_1)}\right)\right. \\ & \,\left.+ \frac{b_{02}(x_1)}{\alpha(q_1)\alpha(q_2)} + \frac{b_{01}(x_1)}{2} f(x_1) (\bs{q}_{12}\cdot \hat{\bs{x}}_1)\left(\frac{(\bs{q}_1\cdot\hat{\bs{x}}_1)}{q_1^2\alpha(q_2)}+\frac{(\bs{q}_2\cdot\hat{\bs{x}}_1)}{q_2^2\alpha(q_1)} \right) \right].
\end{aligned}
\end{equation}
For models of the scale dependent biases, we follow the expressions in \cite{tellarini2016} and write the Eulerian biases (we drop the explicit dependence on comoving distance)
\begin{subequations}\label{eq:png_bias}
\begin{align}
b_{01} &= b^L_{01},
\\
b_{11} &= b^L_{01} + b^L_{11},
\\
b_{02} &= b^L_{02},
\end{align}
\end{subequations}
in terms of Lagrangian biases which, if one assumes a universal mass function, are given by
\begin{subequations}\label{eq:png_bias2}
\begin{align}
b^L_{01} &= 2 \, \delta_c \, \fNL \, b^L_{10},
\\
b^L_{11} &= 2 \, \fNL (\delta_c \, b^L_{20}- b^L_{10}),
\\
b^L_{02} &= 4 \, \fNL^2 (\delta_c \, b^L_{20}-2\,  b^L_{10}).
\end{align}
\end{subequations}
Above, $\delta_c=1.686$ is the critical density for spherical collapse in an Einstein-de Sitter universe. The Lagrangian biases in \eqref{eq:png_bias} can also be related to the Eulerian biases,
\begin{subequations}
\begin{align}
b^L_{10} &= b_{10}-1,
\\
b^L_{20} &= b_{20} - (8/21) \, b^L_{10}.
\end{align}
\end{subequations}
such that all bias parameters can simply be written in terms of $b_{10}$ and $b_{20}$.

The full redshift space kernels, at each order in perturbation theory, are
\begin{equation}
    Z^{(1)} = Z^{(1)}_{\rm G} + Z^{(1)}_{\rm N,NG},
\end{equation}
and
\begin{equation}
    Z^{(2)} = Z^{(2)}_{\rm G} + Z^{(2)}_{\rm N,NG},
\end{equation}
and so the total PNG contribution to the bispectrum can be calculated using Equation~\eqref{eq:full_local_bk}.

\bibliographystyle{jhep}
\bibliography{main}

\end{document}